\documentstyle[aasms4,10pt,epsf]{article}
\def\lsim{\, \lower2truept\hbox{${< \atop\hbox{\raise4truept\hbox{$\sim$}}}$}\,}
\def\gsim{\, \lower2truept\hbox{${> \atop\hbox{\raise4truept\hbox{$\sim$}}}$}\,}

\def\oneskip{\vskip\baselineskip}
\def\sun{\odot}

\def\puncspace{\ifmmode\,\else{\ifcat.\C{\if.\C\else\if,\C\else\if?\C\else%
\if:\C\else\if;\C\else\if-\C\else\if)\C\else\if/\C\else\if]\C\else\if'\C%
\else\space\fi\fi\fi\fi\fi\fi\fi\fi\fi\fi}%
\else\if\empty\C\else\if\space\C\else\space\fi\fi\fi}\fi}
\def\SP{\let\\=\empty\futurelet\C\puncspace}
\def\micron{~$\mu$m\SP}

\def\etal{~~{\it et al.}\SP}

\begin{document}

\title{Models for Multiband IR Surveys}
\author{Cong Xu,  Carol J. Lonsdale, David L. Shupe, JoAnn O'Linger, 
Frank Masci}
\affil{Infrared Processing and Analysis Center, Jet Propulsion Laboratory, 
Caltech 100-22, Pasadena, CA 91125}
\received{Feb. 4, 2001}
\accepted{}

\begin{abstract}

Empirical 'backward' galaxy evolution models for IR-bright 
galaxies are constrained using multiband IR surveys.
A new Monte-Carlo algorithm is developed for this task. It
exploits a large library of realistic Spectral Energy
Distributions (SEDs) of 837 local IR galaxies (IRAS 25$\mu m$
selected) from the UV (1000{\AA}) to the radio (20cm), including
ISO-measured 3--13$\mu m$ unidentified broad features (UIBs).  The
basic assumption is that the local correlation between SEDs and
Mid-Infrared (MIR) luminosities can be applied to earlier epochs of
the Universe, an assumption which will be strongly tested by SIRTF.  
By attaching an SED appropriately drawn from the SED
library to every source predicted by a given model, the algorithm
enables simultaneous comparisons with multiple surveys in a wide range 
of wavebands. Three populations of IR
sources are considered in the evolution models.
These include (1) starburst galaxies, (2) normal late-type
galaxies, and (3) galaxies with AGN. Constrained by data from the
literature, our best-fit model (`Peak Model') predicts that since
z=1.5 the population of starburst galaxies undergoes a very strong
luminosity evolution ($L=L_0\times (1+z)^{4.2}$) and also  strong
density evolution ($\rho=\rho_0\times (1+z)^{2}$), the normal
late-type galaxy population undergoes a passive luminosity evolution
($L=L_0\times (1+z)^{1.5}$), and the galaxies with an AGN undergo a
pure luminosity evolution similar to that of optical QSOs
($L=L_0\times (1+z)^{3.5}$). Prior at z$\geq$1.5 all evolution rates
drop as $(1+z)^{-3}$. 
The luminosity evolution results in evolution of SEDs of 
IR-bright sources because of the luminosity dependence of the
SEDs. Predictions for number counts, confusion limits,
redshift distributions, and color-color diagrams are made for
multiband surveys using the upcoming SIRTF satellite.  A
$\Lambda$-Cosmology ($\Omega_\Lambda =0.7$, $\Omega_m = 0.3$, H$_0$=75
km sec$^{-1}$ Mpc$^{-1}$) is assumed throughout the paper.

\end{abstract}

\keywords{galaxies: starburst -- Seyfert -- luminosity function; 
infrared: galaxies}

\section{Introduction}

The first sign of cosmic evolution among infrared (IR) galaxies was
detected by Hacking, Condon and Houck (1987) in the IRAS
60$\mu m$ deep survey (Hacking and Houck 1987). This was
subsequently confirmed by later studies of IRAS galaxy populations
(Franceschini et al. 1988; Lonsdale and Hacking 1989; Lonsdale et al. 1990;
Rowan-Robinson et al. 1990; Saunders et al. 1990; Yahil et al. 1991;
Gregorich et al. 1995; Pearson and Rowan-Robinson 1996;
Bertin et al. 1997). Recently, deep MIR-FIR surveys have been carried
out using the Infrared Space Observatory (ISO) (Kessler et al. 1996). 
These include ISOCAM surveys at 15$\mu m$, 12$\mu m$ and at 6.7$\mu m$
(see Elbaz et al. 1998b for a summary of these observations),
ISOPHOT surveys at 90$\mu m$ (Oliver et al. 2000, Efstathiou et
al. 2000a) and at 175$\mu m$
(Kawara et al. 1998; Puget et al. 1999; Dole et al. 2001).
The results from these 
surveys (Aussel et al. 1999; Puget et al. 1999; Dole et al. 2001;
Clements et al. 1999; Elbaz et al. 1999; Serjeant et al. 2000;
Xu 2000)
indicate strong cosmic evolution in the population of
infrared-emitting galaxies, confirming the earlier results based
on smaller samples and less sophisticated analyses
(e.g. Rowan-Robinson et al. 1997; Kawara et al. 1998).
This is consistent with the results of SCUBA surveys
(Hughes et al. 1998; Barger et al. 1998; Blain et al. 1999) 
and with the scenario hinted at by the newly
discovered Cosmic IR Background (CIB) (Puget et al. 1996; Hauser et al. 
1998; Dwek et al. 1998; Fixsen et al. 1998), while challenging the
results from UV/optical surveys in the sense
that substantially more (i.e. a factor of 3 -- 5) star formation 
in the earlier Universe is required to match the IR/submm 
counts and the CIB (see, e.g. Rowan-Robinson et al. 1997)
compared to that derived from 
the UV/optical surveys (Madau et al. 1998; Pozzetti et al. 1998,
Steidel et al. 1999).
The reason for this discrepancy is attributed to dust extinction
which may hide much of the star formation in the early Universe
from the UV/optical surveys (see Lonsdale 2000 for a review).

Compared to the UV and optical surveys, 
the infrared surveys are superior in their insensitivity to 
dust extinction, but are inferior in angular resolution
(a few arcsecond compared to the sub-arcsecond resolution of
optical surveys).
This not only limits the IR surveys by confusion, but also
makes the study of the IR morphology of faint IR sources 
impossible. In order to reveal the true nature of faint
IR sources, identifications in other bands, especially in optical
and NIR bands where the sources can be resolved easily with current
instruments, are usually needed. The multi-band studies
(including ISOCAM 15$\mu m$ and 6.7$\mu m$ surveys)
of the Hubble Deep Field (HDF) by Rowan-Robinson et al. (1997) and
Aussel et al. (1999), and of the Canada-France
Redshift Survey (CFRS) field by Flores et al. (1999)
suggest that, compared to their optical counterparts,
the ISOCAM sources have significantly redder 
(I-K) colors (Flores et al. 1999), and 
are much more likely to be in the interacting/merging systems.
On the other hand, such studies are necessarily confined
to IR sources that are relatively bright 
in the optical (e.g. $I \leq 22.5$ mag, 
Flores et al. 1999), while many IR bright galaxies are
optically faint due to heavy dust extinction.

IRAS studies showed that galaxies of different nature in the local universe have
distinct IR spectral energy distributions (SEDs). 
Galaxies with bright active galactic nuclei (AGN) usually have
significantly lower 
$f_{60\mu}/f_{25\mu}$ ratios (de Grijp\etal 1985;
Fang et al. 1998) than other galaxies. Interacting/starburst galaxies
such as M82 have systematically high $f_{60\mu}/f_{100\mu}$ 
and $f_{25\mu}/f_{12\mu}$ ratios than normal galaxies such as the
Milky Way (e.g. Helou 1986). In principle, these different
characteristics in the IR SEDs for different populations of galaxies
should facilitate a tool for identifications of IR galaxies when
multi-band IR surveys are available, independent of the optical
identifications. For ISO surveys this may not be very relevant
because the ISOPHOT FIR surveys do
not match the ISOCAM MIR surveys in depth due to severe
degradation of the sensitivity of ISOPHOT detectors. However, when the
Space Infrared Telescope Facility (SIRTF) is launched in mid 2002,
simultaneous deep surveys in seven MIR-FIR bands (3.6$\mu m$, 4.5$\mu m$,
5.8$\mu m$, 8.0$\mu m$, 24$\mu m$, 70$\mu m$, and 160$\mu m$)
will be possible (Bicay et al. 1999).  These will include the 
Guaranteed Time Observer programs with MIPS and IRAC 
(sirtf.caltech.edu/ROC/Titles\_abstracts.html), 
the large-area Legacy survey SWIRE (Lonsdale {\it et al.} 2001; 
and the very deep Legacy survey GOODS (Dickinson et al. 2001).

Significant K corrections will occur in the 
observed SEDs of faint sources in deep IR surveys. In particular,
in the rest frame
wavelength range between 3--20\micron there are several broad band
features (see Puget and L\'eger 1989 for a review),
often referred to the Unidentified Infrared Bands (UIBs), which 
are ubiquitously
present in the MIR spectra of local galaxies with equivalent widths up to
several microns (Helou et al. 2000), with the exception of type 1 Seyferts
(Clavel et al. 2000).  If these features are also
present in the SEDs of high redshift galaxies, substantial
K-corrections will result in when
any of the features redshifts in or out of the band pass of an IR
filter. These effects may indeed be beneficial
rather than annoying, for they may facilitate 
IR photometric redshift techniques.

Xu et al. (1998, hereafter Paper I) studied the effect of
K-corrections due to UIBs on number counts of MIR surveys.  In that
work, a three-component model, with empirically determined MIR SED
templates of (1) a cirrus/PDR component (2) a starburst component and
(3) an AGN component, is developed for infrared (3--120\micron) SEDs
of galaxies.  The model is then applied to a complete IRAS 25\micron
selected sample of 1406 local galaxies ($z \leq 0.1$; Shupe et
al. 1998).  Results based on these 1406 spectra show that the MIR
emission features cause significant effects on the redshift dependence
of the K-corrections, which in turn affect deep counts and redshift
distributions in MIR surveys.  In a subsequent paper, Xu (2000,
hereafter Paper II) found that indeed the sharp peak at about 0.4mJy
in the Euclidean normalized differential counts at 15$\mu m$ (Elbaz et
al. 1999) can be explained by the effects of UIBs, together with
an evolution rate significantly stronger 
than derived in previous IRAS studies,
eliminating the need for a hypothetical 'new population' (Elbaz 1998a).

In this paper, we expand the models in Paper I in several aspects.
\begin{description}
\item{(1)}
First of all, the analytical algorithm of the number count model which
includes a proper treatment of the K-correction (Eqs (23),
(24) and (25) in Paper I) is replaced by a Monte-Carlo algorithm,
in which every source (galaxy) in a volume of given redshift and 
in a given luminosity bin is assigned an SED appropriately
selected from the SED library (837 SEDs). The source's flux densities
in different bands are then calculated by convolving the
redshifted SED with the band passes of filters. In this way, we effectively
simulate a virtual sky for a given evolution model. This not only
enables the simultaneous comparison of counts in different
bands, but also preserves 
the correlations between flux densities of different bands.
This latter feature allowing us to 
predict color-color diagrams of different
populations as a function of redshift, facilitating the exploration
of photometric redshift indicators. 
\item{(2)}
In the 'backward evolution' model, instead of treating all 
IR sources as a single population, in this work 
they are separated into three populations,
in a similar spirit as in the model of Franceschini et al. (1988;
see also Roche and Eales 1999):
(1) normal late-type galaxies, (2) interacting/starburst galaxies,
and (3) galaxies with AGNs. These different populations are
assumed to have different cosmic evolution rates.
\item{(3)}
The wavelength coverage of our SED library is expanded from 3---120$\mu m$ to 
1000{\AA} --- 20cm. This is done by collecting from the literature the
optical/NIR (B, J, H, Ks bands) magnitudes and the radio continuum (20cm)
flux densities for galaxies in our SED sample, and by
extrapolating from IRAS 60$\mu m$ and 100$\mu m$ bands
to submm bands using empirically determined correlations.
\end{description}

The goal of this paper is to provide a set of comprehensive 'backward
evolution' models (a category
of galaxy evolution models in which number densities and other
properties, e.g. luminosities in different bands, of local 
galaxies are evolved 'backward' in time from the present --- i.e. with increasing redshift
--- according to some parametric prescriptions,
see Lonsdale 2000 for a review) for future multi-band surveys,
in particular those to be conducted with SIRTF. The model parameters
will be constrained by observations available in the 
literature. This not only includes the ISO deep surveys, but also
the optical/NIR surveys, SCUBA surveys and radio deep surveys, 
because our SEDs now cover all of these wavebands.
Constraints derived from the Cosmic IR Background (CIB) will also
be incorporated. The strength of models being constrained by
such wide range of data has already been demonstrated in several 
previous papers (e.g. Blain et al. 1999; Trentham, Blain and 
Goldader 1998; Adelberger and Steidel 2000; Rowan-Robinson 2001).

The model focuses on IR-bright galaxies, and therefore is not expected to
match observed number counts in any band for which there is a substantial 
contribution from IR-quiet populations, eg. the K band,
 because E/S0 populations
are missing from our model, and the bright radio counts, 
which are dominated by radio galaxies.
 
Throughout the paper, the cosmology model specified by
the following parameters is adopted: H$_0$=75 km sec$^{-1}$ Mpc$^{-1}$,
$\Omega_m = 0.3$, $\Omega_\Lambda = 0.7$.

\section{SED Library}
As in Paper I, our SED sample is drawn from 
the IRAS 25\micron selected sample (complete down to
$f_{25\mu}=0.25$~Jy) by Shupe et al. (1998), 
and contains 1455 galaxies, 1406 of them with redshifts $\leq 0.1$.
As pointed out by Spinoglio et al. (1995),
the MIR luminosities correlate well with the bolometric
luminosities. Therefore MIR selected
samples, such as ours, have fair representations of different 
populations of IR sources. However E/S0s, which are $\sim 20\%$
of optically selected galaxy samples but mostly undetected
by IRAS, are not included in our SED sample.

In Paper I, a three-component (cirrus/PDR, starburst, and AGN) MIR-SED
model is applied to these galaxies, predicting an SED from
3--120\micron for each of them.  In order to expand the SEDs to the
optical and NIR bands, B (4400{\AA}), J (1.0\micron), H (1.6\micron),
and Ks (2.2\micron), we searched the literature. B
magnitudes of 1339 galaxies were found in the NASA/IPAC Extragalactic
Database (NED). The NIR magnitudes are taken mainly from the 2MASS
Second Incremental Data Release via the IRSA facility\footnote{The
NASA/IPAC InfraRed Science Archive (IRSA) is a NASA project focused on
providing software and Internet services to facilitate astronomical
discoveries, support the production of new astronomical data products,
and to plan future observations utilizing the data archives from
infrared astrophysics missions supported at Infrared Processing
and Analysis Center (IPAC).}
where J, H, and Ks magnitudes of 790 galaxies were found.
In addition, NIR magnitudes of 413 galaxies in our sample are 
given in Spinoglio et al. (1995), 244 of which are overlapped with the 2MASS
matchs. Whenever NIR magnitudes are available from both
2MASS and Spinoglio et al. (1995), 2MASS data take precedence.
Altogether, J H and Ks magnitudes are found for 959 galaxies in our sample.
The radio continuum flux densities at 20 cm, $S_{1.4GHz}$, were
searched for in both the NVSS (Condon et al. 1998)
 and FIRST (Becker et al. 1995) catalogs of NRAO. Among 1406 galaxies 
in our sample 1170 are found in one of these two surveys 
(the rest are in the sky area not visible by VLA).
It is found that 854 galaxies in our sample have B, J, H, Ks 
magnitudes and radio continuum flux at 1.4GHz. After excluding 
17 galaxies which were undetected by IRAS both in the 60\micron and
100\micron bands (whose IR SEDs are highly uncertain), we select
a final SED sample of 837 galaxies. Note that the 569 galaxies 
in the original sample (1406 galaxies) that do not make it into
the final SED sample are mostly galaxies without NIR magnitudes. This is
mainly due to the fact that data for NIR sources in
a large fraction of the sky has not been releaseded by
the 2MASS survey (the major source of the NIR data) yet.
Since both the 2MASS survey and the VLA
surveys are much deeper than the IRAS survey, 
the NIR magnitudes or the 20 cm flux density is missing for a source
in the 25$\mu m$ selected sample only when the sky region is missing
in the corresponding database. Therefore no bias is introduced into
the final SED sample when sources without NIR or radio fluxes are
excluded.

For each galaxy in the SED sample, the broad-band 
UV-optical-NIR (1000{\AA} -- 4\micron) 
SED is estimated by a spline fit of the fluxes in
B, J, H, Ks bands, altogether with the predicted 4\micron flux density
from the MIR SED model (Paper I). It should be noted that at the
UV wavelengths (1000 --- 4000{\AA}) the predicted fluxes
are extrapolations from the available data, and caution should be
applied when these predictions are used.  This aspect of the model will
be improved in the next paper, using the new UV data that are 
just now becoming available for ULIRGs and other IR-bright galaxies.

The MIR (4 -- 16$\mu m$) SED is determined 
using the MIR SED model developed in Paper I, including a full treatment of the
UIB features (absent in type 1 AGN).  Then the SED in the
wavelength range 16 --- 1200 $\mu m$ is specified by a spline fit
of IRAS data at 25, 60, 100\micron, together with the 16\micron
flux density predicted by the MIR SED model, and the 
170$\mu m$, 240$\mu m$, 450$\mu m$, 850$\mu m$ and 1200$\mu m$ 
flux densities predicted by empirical correlations between 
the given submm band flux and the IRAS 60$\mu m$ and 100$\mu m$
fluxes which are derived
from available submm data collected from the literature
(Appendix).

The radio continuum flux density at 20cm ($S_{1.4GHz}$) is extrapolated to 
6.2cm ($S_{4.8GHz}$) and 2.8cm ($S_{10.2GHz}$) using the mean
spectral indices $\alpha_{20cm/6.2cm}=0.79$ and
$\alpha_{6.2cm/2.8cm}=0.70$, found  for 
Shapley-Ames galaxies (Niklas et al. 1997). 
These radio flux densities are then linked to
the end of the IR/submm SED at 1200$\mu m$ by spline fit.

In Fig.1 we show the SEDs which are binned according to  
population (see the next section)
and the 25$\mu m$ luminosity (Table~1). For each bin,
the mean SED and its 1$\sigma$ dispersion are also
plotted in the corresponding panel in Fig.1, and
the values are listed in Table 2.

As examples, in Fig.2 observational data of twelve well known galaxies
are compared with model SEDs. The data are collected
from the literature, and the sources are
\begin{itemize}
\item broad band optical magnitudes: NED; 
\item NIR magnitudes: 2MASS and Spinoglio et al. (1995);
\item FIR/submm flux densities: IRAS, Benford (1999), Dunne et al. (2000), 
Rigopoulou et al. (1996), Lisenfeld et al. (2000), Carico et al. (1992),
Chini et al. (1986), Andreani and Franceschini (1996), 
Roche and Chandler (1993);
\item radio continuum flux densities: Niklas et al. (1995), Condon 
et al. (1990).
\end{itemize}
The agreements between the data and the model SEDs are remarkably 
good in general.

\section{'Backward Evolution' Model for Multi-band Surveys}
\subsection{Three Populations of IR Emitting Galaxies and Their LLFs}
In Paper I and Paper II, it is assumed that all 
IR sources evolve as a single population.  Here we improve on that
formulation by adopting a model in which IR sources
can be separated into three populations,
in a similar spirit as in the model of Franceschini et al. (1988;
see also Roche and Eales 1999):
(1) normal late-type galaxies, (2) interacting/starburst galaxies,
and (3) galaxies with AGNs. 

To enable these different galaxy populations to evolve at different rates,
the model requires a local luminosity function (LLF) for each component.  We
began with the 25$\mu m$ flux density-limited sample of 1455 galaxies
of Shupe et al. (1998).  Although classifications of many of the
galaxies as AGNs, normal galaxies, etc. are available in databases
such as NED, to treat the sample in a more uniform way, we
use IR-color-based criteria to divide our sample into
different populations (see Paper I and Fang et al. 1998
for a discussion).  
We chose the following IRAS color boundaries:
\begin{itemize}
\item   ``AGNs":  $f_{60\mu m}/f_{25\mu m} \leq 5 $
\item   ``Starbursts": $f_{60\mu m}/f_{25\mu m} > 5 $ and 
$f_{100\mu m}/f_{60\mu m} < 2 $. 
\item   ``Normals": $f_{60\mu m}/f_{25\mu m} > 5 $ and 
$f_{100\mu m}/f_{60\mu m} \geq 2 $.
\end{itemize}

This division results in 356 galaxies classified as ``AGNs",
456 galaxies as ``Normal", and 643 galaxies as
``Starburst".  We note that our color-selection method for AGN will not
be appropriate for heavily obscured objects in which even the 
$f_{60\mu m}/f_{25\mu m}$ color may be affected by reddening; instead any such
IR-red AGN will be found in the `starburst' sample. 
Also, for many AGNs as defined above, much of the IR radiation can
be due to the emission of dust heated by stars in the host galaxy
in addition to the dust emission associated with the AGN.
For example, according to Eq(1) of Paper I, an IR source of 
$f_{60\mu m}/f_{25\mu m} = 5 $ (an `AGN' by the above definition)
has a half of its 25$\mu m$ emission from dust heated by stars.
In reality, Seyfert galaxies such as NGC~4945 (Spoon et al. 
2000) can have the IR emission predominantly powered by the nuclear starburst
rather than by the AGN.  This explains why the SEDs of some sources
in the AGN subsample of our SED library show significant
broad band MIR emission features (Fig.1a) which should be absent
in a typical type 1 AGN SED (e.g. Fig.3 of Paper I). At the same time,
as shown in Fig.2, the UIB features are indeed absent
in the model SEDs of AGNs such as Mrk~231 amd NGC~7479, consistent
with the fact that the emission in these sources is dominated
by the AGN. It should also be noted that in our SED model (Paper I), 
we do not distinguish type 1 and type 2 AGNs, which have 
significantly different MIR SEDs (Clavel et al. 2000).
Many type 2 AGNs have strong UIB features even when their
IR emission may be predominantly from dust associated with AGN,
because an edge-on torus may heavily extinguish the MIR part of
the AGN-associated emission and therefore
the detected MIR emission is mostly from dust in the ISM 
of the host galaxy (Clavel et al. 2000).
 
Luminosity functions were then computed for each of
these populations according to
the maximum-likelihood method described in Yahil et al. (1991) and
used in Shupe et al. (1998).  This method calculates the shape of
a parametric luminosity function described by the parameters
$\alpha$, $\beta$ and $L_*$ independent of density variations.

With the shape parameters in hand, the
normalization of each luminosity function must be estimated by other
methods.  Since the normalization of the total 25 micron luminosity
function was estimated in Shupe et al. (1998), the normalizations
of the three population LFs are chosen so that the the number of
galaxies implied by the sum of the component LFs is about the same
as the total LF.  
The difference between the total LF and summed LF is also constrained
to be less than a few percent at all luminosities.  We have adjusted
the normalizations of the component LFs to satisfy these criteria.
These relative weightings of the component LFs can be adjusted by 
ten to twenty percent while still satisfying the criteria, but cannot
be made vastly different from the nominal values.

A plot of the component luminosity functions, the sum of the component
LFs, and the total LF is shown in Fig.3.  The
computed parameters for each population are given in Table 3.
Note that the LFs are calculated using the whole sample (1455
galaxies) of Shupe et al. (1998).
Identical results are obtained when the sample is confined to the 1406
galaxies with $z<0.1$.

\subsection{Monte-Carlo Simulation of Multi-band Surveys}
In this subsection, we develop the algorithm to model coherently
the number counts in different bands and the color-color diagrams for
multi-band surveys.

In a flat $\Lambda$-Universe (i.e. $\Omega = \Omega_m +\Omega_\Lambda
=1$, $\Omega_\Lambda \neq 0$), which is adopted in this work,
the co-moving volume is
\begin{equation}
V  =  {A\over 3}\; D_M^3
\end{equation}
where A is the sky coverage in steradians, and
$D_M$ is the proper motion distance (Carroll et al. 1992)\footnote{
The $D_A$, the angular diameter distance in Section 4.1 of Paper I, should have
been called $D_M$, the proper motion distance, too.}:
\begin{equation} 
D_M = {c\over H_0} \int_0^{z1} [(1+z)^2 (1+\Omega_m z) - 
z(2+z)\Omega_\Lambda]^{-1/2} d\; z
\end{equation} 

The predicted number of sources from a given population, in a given
redshift interval $[z-0.5\delta z, z+0.5\delta z]$ and in
a given 25$\mu m$ luminosity
interval $[L-0.5\delta L, L+0.5\delta L]$, is then
\begin{equation} 
\delta N_i(L,z) = \rho'_i(L,z) {dV\over dz} \delta L\,  \delta z \label{eq:N}
\end{equation} 
where $\rho'_i$ is the luminosity function of population $i$ 
($i=1$ -- normal late type galaxies, $i=2$ -- starburst galaxies, 
and $i=3$ -- galaxies with AGNs): 
\begin{equation}
\rho'_i (L,z) = G_i(z)\, \rho_i \left( {L\over F_i(z)} \right);
\end{equation}
where $\rho_i$ is the local 25\micron luminosity function (luminosity function
at z=0) of population $i$ (Section 3.1), and 
$G_i(z)$ and $F_i(z)$ are the density evolution function and 
the luminosity evolution function of population $i$, respectively.

For each of the $\delta N_i(L,z)$ sources predicted using
Eq(\ref{eq:N}), an SED is randomly selected from a proper SED bin in
the SED library (Section 2) and is assigned to the source. 
The SED library is binned according to
(1) the population, and (2) the 25$\mu m$ luminosity (Table 1).
For an individual source the 
flux density in a given band can then be determined as
follows:
\begin{equation}
f_{band} ={1\over 4\pi D_L^2} \times 
{L_{25\mu m}/25\mu m\over S(25\mu m)} 
\int_{\lambda_1}^{\lambda_2} S({\lambda\over 1+z}) R_{band} (\lambda) d\lambda
\end{equation}
where 
\begin{equation}
D_L = (1+z) \times D_M
\end{equation}
is the luminosity distance (Carroll et al. 1992), $L_{25\mu m}=\nu
L_\nu (25\mu m) = \lambda L_\lambda (25\mu m)$ is the monochromatic
luminosity at 25$\mu m$, $R_{band} (\lambda)$
is the bandpass of the given band, and $S(\lambda)$ is the flux density 
distribution of the SED in question.   Due to the dependence of SED shape
on luminosity for IR-bright galaxies (the $f_{60\mu}/f_{100\mu}$ color
increases with $L_{IR}$), this approach empirically results
in color evolution accompanying luminosity evolution.

When the evolution functions in Eq(4) are specified, we can 
predict counts in different IR bands, as well as
contributions from IR galaxies to counts in other wavebands.
As a test, local luminosity functions in the IRAS 60$\mu m$ band
(Fig.4),
in the SCUBA 850 $\mu m$ band (Fig.5), and in the IRAM
1250$\mu m$ band (Fig.6) are calculated via model simulations 
(for sources of $z<0.1$) specified by $G_i(z)=1$ and $F_i(z)=1$.
Good agreement with the IRAS 60$\mu m$ luminosity function
of Saunders et al. (1990), with the 850$\mu m$ luminosity function
of Dunne et al. (2000), and with the 1250$\mu m$ luminosity
function of Franceschini et al. (1998) is found.

\subsection{Evolution Models}
As in Paper I and Paper II,
the following (power-law) function forms are  adopted for 
the luminosity evolution functions $F_i(z)$ and the density evolution
functions $G_i(z)$:
\begin{eqnarray}
F_i(z) & = & (1+z)^{u_i} \;\; (z \leq z_{1}) \nonumber \\ 
     & = & (1+z)^{v_i} \;\; (z_{1} < z \leq z_{0})
\end{eqnarray}
\begin{eqnarray}
G_i(z) & = & (1+z)^{p_i} \;\; (z \leq z_{1}) \nonumber \\ 
     & = & (1+z)^{q_i} \;\; (z_{1} < z \leq z_{0})
\end{eqnarray}
where $z_0$ is the redshift when the galaxy formation started,
and $z_1$ is the so called `peak' redshift, where the evolution reaches
a peak.  Here we explore two kinds of models, 
the first characterized by a steady
increase in evolution from $z_0$ to $z_1$ at power law rates $v_i$
and $q_i$ followed by a strong decline to the present day with power law
rates $u_i$ and $p_i$, and the second having a plateau between the formation
and peak epochs, $z_0>z>z_1$.

Throughout the paper we adopt $z_0=7$ and $z_1=1.5$.  Our results 
(predictions for number counts and the CIB) are not sensitive
to $z_0$ so long as it is $>5$.
Optical surveys (e.g. Lilly et al. 1996; Madau et al. 1996;
Connolly et al. 1997)
show that star formation rate in galaxies peaks between
$z=1$ and $z=2$.   The deep ISOCAM counts (e.g. Elbaz et al. 1999)
are consistent with a rapidly increasing star formation rate
going back at least to $z\geq 1.5 $ (Paper II).
The choice of 1.5 for the peak redshift is driven by the 
deep ISO 15$\mu m$ data (see Fig.7).   It is still controversial
whether the star formation rate indeed decreases at $z>z_1$, beyond 
the peak, or flattens (e.g Steidel et al. 1999; Blain et al. 1999). 

Other parameters are $u_i,v_i,p_i,q_i$ (i=1, 2, 3).
In order to reduce further the parameter space, we assume:
\begin{description}
\item{(1)} For normal late type galaxies (population 1): $u_1=1.5$,
$p_1=0$.  This corresponds to a 
pure passive luminosity evolution before the turn-over redshift.
These  galaxies are a major constituent of K-band extragalactic
source counts, which may be consistent with passive evolution
models (Gardener et al. 1997).
\item{(2)} For galaxies with AGNs (population 3):
$u_3=3.5$, $p_3=0$. Here the assumption of 
the pure luminosity evolution is based on the studies of the evolution
of QSOs in the literature (e.g. Boyle et al. 1988; Pei 1995; La Franca
et al. 2000).
Using a maximum-likelihood technique Boyle et al. (1988) found that 
pure luminosity evolution models of form
$L(z) \propto L_0\times (1+z)^\gamma$ ($\gamma = 3.2\pm 0.1$) adequately 
describe the evolution of bright ($M_B < -23$)
low reshift ($z< 2.2$) QSOs. A similar result was found by
Pei (1995), with $\gamma$ in the range of 3.2 --- 3.9.
Recently, La Franca et al. (2000) found that results from 
an ISOCAM 15$\mu m$ survey of Type 1 AGNs in the ELAIS fields
are consistent with a pure luminosity evolution model with
$L(z) \propto L_0\times (1+z)^{3.4}$.
\item{(3)} The evolution indices $u_2$ and $p_2$, which specify
the luminosity and density evolution rate of
starburst galaxies before $z=1.5$, will be the free parameters.
Given the results of Paper II on the evolution of deep ISOCAM counts,
it is expected that at $z<1.5$ 
the starburst galaxies will have a stronger
evolution rate  than what has been 
assumed for normal late-type galaxies (assumption
(1)) and for galaxy with AGNs (assumption (2)), consistent with
the preliminary identifications of faint
ISOCAM sources (Flores et al. 1999).
\item{(4)} Again for the sake of simplicity, we assume that
all above mentioned evolution rates will have the same behavior
after $z=z_1$, namely
$v_1=v_2=v_3=q_2$. This is because beyond z$\simeq 1.5$
(the highest redshift detected for ISO galaxies), we know very
little about the population of IR galaxies. 
\end{description}

\section{Comparisons with Available Surveys and Constraints on
Evolution Parameters}
\subsection{Comparisons with ISOCAM 15$\mu m$ Band Surveys: 
Constraints on Evolution of $z<1.5$}
In what follows, we will compare our simulations
with surveys in different bands found in
the literature. The evolution models considered in this paper
are listed in Table 4. 

We start with the surveys in ISOCAM 15$\mu m$ band
(Elbaz et al. 1999; Serjeant et al. 2000), where the deepest
and the most comprehensive ISO surveys have been conducted.
Due to significant effects caused by the UIBs, certain
features in the MIR counts can help to constrain the rate
of luminosity evolution and that of density evolution 
separately (Paper I). If the sharp peak at 
$f_{15\mu m} \simeq 0.4 mJy$ in the Euclidean normalized
differential counts is indeed due to the UIBs
in 6 --- 8 $\mu m$ (Paper II), which are redshifted into the 15$\mu m$ band
when $z\sim 1$, then a 15$\mu m$ luminosity of 
$\sim 10^{11} L_\sun$ ($\nu L_{nu}$ at 15$\mu m$)
could be inferred for a typical z=1 ISOCAM galaxy. 
This imposes a strong constraint to
the luminosity evolution rate of the major population of the ISOCAM
sources, which under our assumptions (Section 3.3) are the starburst 
galaxies, at $z \lsim 1$.

In Fig.7a and Fig.7b, we compare the simulations of
two evolution models, one has $p_2=2$, $u_2=4.2$, $z_1=1.5$ and
$v_1=v_2=v_3=q_2=-3$ 
(Model 1, `Peak Model') and the other $p_2=2$, $u_2=4.2$,
$z_1=1.5$ and $v_1=v_2=v_3=q_2=0$  (Model 2, `Flat Model') with 
the number counts of ISOCAM 15$\mu m$ data surveys  
(Elbaz et al. 1999; Serjeant et al. 2000). 
Both models fit the ISO data very well. Namely,
although the `Flat Model' predicts about a factor of 2 more sources 
at $f_{15\mu m}=0.01$mJy,
there is little difference between the two models
above the sensitivity limit of ISOCAM surveys ($~0.1 mJy$).
At $~0.1 mJy$
the 15$\mu m$ counts are very insensitive to 
galaxy evolution at $z>1.5$.

In Fig.7c, 7d simulations of two other evolution models,
Model 3 ($p_2=1$,
$u_2=5$, $z_1=1.5$, $v_1=v_2=v_3=q_2=-3$) and Model 4 ($p_2=3$, $u_2=3.5$, $z_1=1.5$, $v_1=v_2=v_3=q_2=-3$) 
are also plotted. The former assumes more luminosity
evolution ($u_2=5$) and less density evolution ($p_2=1$) for
the starbursts at $z<1.5$. It gives a slightly
less good fit to the 15$\mu m$
data (Fig.7c), predicting a shallower peak at slightly brighter flux level
than that of the data. The latter assumes less luminosity
evolution ($u_2=3.5$) but more density evolution ($p_2=3$) for
the starbursts before $z\leq1.5$. It fits the 15$\mu m$
data very well (Fig.7d). 

In Fig.8, we compare
the redshift distributions predicted by the `Peak Model'
model and by the `Flat Model' with the data.
Both the data of Aussel et al. (1999) and
of Flores et al. (1999) are assumed to be complete
at the 50\% level (i.e. half of the sources are missing from
the two plots due to lack of redshifts). 
By definition (Table 4), 
the `Peak Model' and the `Flat Model' differ only at $z>1.5$.
For the HDF-North survey, the model predicts a slightly higher
median z ($\sim 1$) compared to the median of the data ($z \sim 0.7$).
However, since the redshift data are not complete and the high 
redshifts are more likely to be missing (more difficult to measure),
and since the HDF-North is such a tiny field that any cluster at a given
redshift (e.g. at $z \sim 0.7$) can affect the redshift distribution
significantly, we feel this discrepancy is not inconsistent with
our models.
The CFRS survey is shallower (and wider) than the HDF-North survey, 
but its redshift distribution is more skewed toward the high z 
end, supporting our argument that the z distribution of the
ISOCAM sources in the HDF-North field might not be representative
for $f_{15\mu m}>0.1$ sources. The `Peak Model' predicts
a median redshift of 0.75 for the CFRS survey, very close to
that of the data.

\subsection{Comparisons with the CIB and the SCUBA 850$\mu m$ Band Surveys:
Constraints on Evolution of $z>1.5$}
Because of the negative K-correction in the submm bands,  
the best constraints on galaxy evolution beyond $z\sim 2$ come
from the CIB and the submm counts. 
Our model predictions for the CIB are derived by summering up
the flux densities of all sources in a very deep simulation
($f_{24\mu m} \geq 10^{-9} mJy$), for the band in question. 
In Fig.9 we compare
the CIB predicted by four different models to the observations.
The solid curve is obtained by the `Peak Model', which
fits the CIB very well (except for the 60$\mu m$ point of
Finkbeiner et al. 2000). 
The dotted curve is the result of the `Flat Model', which
over predicts the submm CIB substantially (by a factor of 
$\sim 2$). The stronger density evolution model 
in Fig.7c (Model 4, dot-dashed curve in Fig.9)
over-predicts the CIB by $\sim$ 30 --- 50\%.
In particular, it marginally violates the upper limits
set by the TeV gamma-ray observations (Stanev and Franceschini
1998) in the MIR. Another model (Model 5)
which is otherwise same 
as the `Peak Model' except for a less steep drop
($v_1=v_2=v_3=q_2=-1.5$ instead of 
$v_1=v_2=v_3=q_2=-3$) after z=1.5, is also plotted in Fig.9
(dashed curve).  It slightly over-predicts the CIB 
around 300$\mu m$.

In Fig.10 we compare results from simulations of
the `Peak Model' and of the `Flat Model' to the 850$\mu m$
SCUBA counts. As in the case of the CIB comparisons, the 
`Peak Model' fits the data very well, while the 
predictions of the `Flat Model' are significantly higher than
the data.

\subsection{Comparisons between Best-fit Model and 
the Surveys in Other Bands}

The agreements between the predictions of our best-fit model, 
the `Peak Model', and the data from
the ISO surveys at 90$\mu m$ (Fig.11) and at 175$\mu m$ (Fig.12)
are very good. 
Note that near the faint
ends of the 175$\mu m$ counts, incompleteness
at level of $\gsim 50$\% is expected (Dole et al. 2000). Corrections
for this incompleteness in those data will make the agreement 
between our model predictions and the data even better.

These good agreements may not be very surprising
given that the model plotted here is mostly constrained by fitting
the 15$\mu m$ survey data (Fig.7), and that the three FIR bands
are linked to the 15$\mu m$ band by a SED model that is very
robust (Section 2).   In Fig.13 the predictions of our best-fit model 
are compared with the IRAS surveys at 60$\mu m$ (Fig.13). 
Here the data show a large spread, and
our model is about 10\% higher relative to the 
data at log $f_{60} <$ 0.3 (Jy)
if the Gregorich et al. (1995) data are ignored (they may 
suffer from over-estimated; Bertin {\it et al.} (1997)).  
If real, the small over-prediction 
of these bright counts might be a consequence
of a high local normalization (cf. Lonsdale {\it et al.} 1990).
We will investigate this possibility further in our next paper.

We compare the redshift distribution 
of IRAS 60$\mu m$ sources observed by Oliver et al. (1996)
and that predicted by our best-fit model in Fig.14.
The overall agreement looks quite good. On the other hand, there seems to
be a slight excess of data points in the low z region ($z<0.014$) and 
the opposite in high z ($z>0.033$) region. It is not clear 
whether these are due
to large scale structures (the amplitudes of the deviations
are of the same order as the fluctuations due to the
large scale structures) or they hint that the luminosity evolution rate of
the best-fit model is too strong. It is also not clear whether
the low fraction of high z galaxies could be explained,
at least partially, by 
the incompleteness of the redshift survey ($\sim 10\%$, Oliver et al.
1996). Future deeper surveys will certainly help to
answer these questions.

Finally we present the predicted contributions of IR-bright sources
to the counts in the NIR K band, the optical B band, and the radio continuum
20cm band, and compare them with survey data from the literature.
Again only the predictions
of our best-fit model, the `Peak Model', are considered here.
Euclidean normalized differential counts in K band are plotted
in Fig.15.  Compared to the data points
taken from various K band surveys (Soifer et al. 1994;
Gardner et al. 1996; Bershady et al. 1998; 
Minezaki et al. 1998), the `Peak Model' 
predicts that for K $< 22$, about 50 to 80\% of the 
sources are IR bright, in reasonable agreement
with the fraction of late-type galaxies among K band sources
(Huang et al. 1998). It appears that the model prediction is 
significantly below the counts of K $> 22$, though there are only
two data points there. A tentative
comparison with HST NICMOS H band deep counts (Storrie-Lombardi,
private communication) shows that the `Peak Model' predicts
about 50\% of the NIR counts down to H=26.5 mag.

In Fig.16, we compare predicted contributions from IR sources
to the B band counts. Since the E/S0 galaxies constituent
only $\sim 20\%$
of the optical galaxies (Glazebrook et al. 1995), which is a general result
holding even for a very faint sample down to $m_I \simeq 24.25$ mag
(Driver et al. 1995), IR-bright  sources should dominate the 
B band counts. The discrepancy near the bright end 
($B\sim 15$) could be explained by 
the local density enhancement (local super-cluster).
At the faint end ($B > 25$), 
the `Peak Model' also slightly under-predicts
the counts by $\sim 50\%$. It should be noted that
at these faint B band flux levels, where many sources 
have high redshifts and the B band flux is actually
due to the rest frame UV emission, 
the model predictions suffer large uncertainties
because the UV SEDs used in the calculation are not well constrained
(Section 2). This will be the focus of our next paper.

In the radio continuum 20 cm band, the bright sources ($> 1mJy$)
are mostly due to the early-type radio galaxies (Condon 1984)
which are not IR emitters, while the faint, sub-mJy sources
are mostly late-type galaxies (Condon 1984). This is indeed
what the `Peak Model' predicts (Fig.17): at flux levels
brighter than 1 mJy, the
IR sources contribute less than 10\% of radio counts. 
At faint flux levels ($\sim 0.1 mJy$), they can fully account
for the radio counts.

\section{Predictions for Multi-Band SIRTF Surveys}
\subsection{Number Counts and Confusion limits}
In this section we will make predictions using our best-fit model
(`Peak Model').
We will concentrate on future surveys with SIRTF, which will be
launched in mid 2002. All three MIPS bands (24$\mu m$,
70$\mu m$, 160$\mu m$) and all four IRAC bands (3.6$\mu m$,
4.5$\mu m$, 5.8$\mu m$ and 8$\mu m$) bands are considered.
For the sake of simplicity, we assume all SIRTF 
bandpasses are 10\%
($\lambda/\delta \lambda =10$)\footnote{See
 ``{\it Space Infrared Telescope Facility (SIRTF) ---
Observer's Manual (Version 1.0)}'' 
for details of SIRTF instruments.}. We note that,
since the E/S0 galaxies are not included in the model,
we may significantly underpredict the counts in the IRAC bands,
particularly for the shorter wavelength bands (i.e. 
the 3.6$\mu m$, 4.5$\mu m$, and 5.8$\mu m$ bands).
Therefore, our predictions for the counts and the
confusion limits in these bands should be treated as lower
limits.

In Fig.18 we plot the predicted integral counts for the
three MIPS bands and three IRAC bands (3.6$\mu m$, 5.8$\mu m$ and 8$\mu
m$). In all MIPS bands, the counts are dominated by the 
starburst component (including any heavily obscured AGN).  In the
IRAC bands, normal spirals
and galaxies with AGNs give significant contributions to
counts brighter than 0.3mJy (in the 3.6$\mu m$ and 5.8$\mu m$ bands
they out-number the starbursts). At fainter flux levels ($< 0.1mJy$)
the starburst component dominates the counts in the IRAC bands as well.

Confusion limits shown in Fig.18
have been computed from the predicted number counts,
assuming parameters appropriate for SIRTF. 
The method used is the same as that used by Hacking
and Soifer (1991)--namely, summing (in quadrature) the
contributions from all sources fainter than the limit within
the beam, via numerical integration of Eq. (19) in Hacking
and Houck (1987).  An Airy function computed from the
nominal wavelength is used for the beam.  For wavelengths
greater than 6$\mu m$, a telescope diameter of 85 cm
is used in accordance with the required diffraction-limited
performance of the SIRTF telescope.  For wavelengths
shorter than 6$\mu m$, an effective telescope diameter
is used to make the beamsize larger.  Specifically,
an effective diameter of 82 cm is used for 5.8$\mu m$,
and 51 cm is used for 3.6$\mu m$.  As noted above, the
confusion limits for the short wavelength IRAC bands such
as the 5.8$\mu m$ and the 3.6$\mu m$
bands should be treated as lower limits because of 
the omission of the contribution from counts due to
E/S0 galaxies. Also for the MIPS 160$\mu m$ band, 
confusion caused by Galactic cirrus may be 
significant, especially in high cirrus regions
(Gautier et al. 1992; Helou and Beichman 1990).

In Fig.19 redshift distributions predicted by the best-fit model
(`Peak Model') and by the `Flat Model' are plotted.
For a 24$\mu m$ survey limited at $f_{24\mu m}=0.055mJy$
(5$\sigma_{conf}$),
the best-fit model predicts a prominent peak at about
z=1. The starburst component overwhelmingly dominates the
counts at $z > 0.5$, while the normal galaxies dominate
the small z end. The contribution from galaxies with AGNs
is never important ($ < 10\%$), although it could be significantly higher
for obscured AGNs classified here as ``starbusrts'' due to their colors.
The shape of the
redshift distribution predicted by the `Flat Model' is
very different, with a second peak at $z\sim 2$, caused
mainly by the prominent UIB features between 6 --- 8$\mu m$
(Paper I). This peak is cut off in the `Peak Model'
because of the steep decrease of the star-formation rate
at z$>$1.5. However, the turn-over redshift $z_1$,
which has been assumed to be 1.5 in all our models presented here,
is poorly constrained because ISO surveys are too shallow whereas
the SCUBA and CIB are more sensitive to sources at $z > 3$.
If in reality $z_1 > 2$, then it will be
revealed by the second peak at $z\sim 2$
in the redshift distribution of SIRTF 24$\mu m$ sources.

\subsection{SIRTF Color-Color Diagrams}

In Fig.20a and Fig.20b we plot the $f_{24\mu m}/f_{8\mu m}$ vs. $f_{70\mu
m}/f_{24\mu m}$ color-color diagrams predicted by
the `Peak Model' and the `Flat Model', respectively. 
Simulations of both models are
confined to 1 deg$^2$ with the following flux limits:
$f_{24\mu m}\geq 0.055mJy$ (5$\sigma_{conf}$) and $f_{70\mu m} \geq 2.6mJy$
(2$\sigma_{conf}$).
Some general trends are visible in these plots.
Galaxies with AGNs
are mostly in the relatively low $f_{70\mu m}/f_{24\mu m}$
ratio region ($\log (f_{70\mu m}/f_{24\mu m}) \lsim 1)$),
while starburst galaxies dominate the high $f_{70\mu m}/f_{24\mu m}$
ratio region ($\log (f_{70\mu m}/f_{24\mu m}) \gsim 1)$).
Normal galaxies, with relatively low luminosities and therefore
seen only with $z<1$, are concentrated in the low 
$f_{24\mu m}/f_{8\mu m}$ ratio region.
It should be pointed out that these trends are closely related
to our definitions of the three populations (Section 2),
which are separated according to the IRAS color ratios
of $f_{60\mu m}/f_{25\mu m}$ and $f_{25\mu m}/f_{12\mu m}$.
While the criteria adopted are based on empirical correlations
(Fang et al. 1998; Paper I), the clear boundaries in the definitions of
the different populations will have artificially
enhanced the trend shown here.

In Fig.21a and Fig.21b we plot the predictions 
by the same two models for the MIPS color-color diagram,
namely $f_{160\mu m}/f_{70\mu m}$ vs. $f_{70\mu
m}/f_{24\mu m}$. The symbols are
the same as in Fig.20. In these plots, in addition to the
flux cut-offs of $f_{24\mu m}\geq 0.055 mJy$ and $f_{70\mu m}\geq 2.6mJy$,
it is also required that $f_{160\mu m}\geq 38mJy$ (2$\sigma_{conf}$).
The simulations cover 10 deg$^2$.  Most high z galaxies (large crosses 
in the plots) are concentrated in the high $f_{160\mu m}/f_{70\mu
m}$ end ($\log (f_{160\mu m}/f_{70\mu m}) \gsim 0.7$), 
due to the K-correction effect associated with
the curvature in the 20---160$\mu m$ wavelength range
in most of the SEDs in our library.

The apparent `tracks' in these figures are a result of the manner in which
an SED is selected for each simulated galaxy from the SED library.
As luminosity increases with increasing z (for luminosity evolution models)
a model galaxy is assigned a random SED from the relevant luminosity bin.
Some bins contain small numbers of SEDs so the `tracks' reflect single SEDs 
randomly assigned to model galaxies at higher and higher z (and therefore L).
The `tracks' can discontinue when model populations `jump' to  
higher SED library luminosity bins.  Although small random scatter 
could be incorporated into these model colors, we have chosen
not to do that so that the figures can be more easily analyzed.

The IRAC color-color diagrams, such as
$f_{4.5\mu m}/f_{3.6\mu m}$ vs. $f_{8\mu m}/f_{5.8\mu m}$
diagram, are affected by the exclusion of
E/S0 galaxies in our current models. Also the NIR spectral features
considered by Simpson and Eisenhardt (1999),
e.g. the 1.6$\mu m$ H$^-$ opacity minimum and the 2.3$\mu m$ CO  
bandhead, are not
considered in our SED model (Section 2), 
therefore the changes in the IRAC
colors due to the K-correction caused by these features are
absent in our model predictions. Improvements in these
aspects will be addressed in our next paper.

\section{Discussion}
\subsection{Comparisons with Previous Models in Literature}
\subsubsection{Comparisons with Models in Paper I and Paper II}
Three new features separate the models presented here from
the previous models in Paper I and Paper II:
\begin{description}
\item{(1)} An approach even more empirical than that used in
Paper I is adopted in this work. Realistic SEDs of local galaxies
are attached to sources of all redshifts. This not only enabled
reliable K-corrections, but also linked the surveys in different
bands coherently. Because of this feature, our model is the
first in the literature that can
predict the correlations and the dispersions
in color-color diagrams for multi-band surveys.
\item{(2)} The MIR-FIR SEDs in the SED sample are extended 
to a much wider range covering from UV (1000{\AA}) to radio
(20cm). This links the model predictions for IR surveys
to the optical and radio continuum surveys. Because of the
wide wavelength coverage, our model can calculate the contributions
from the IR sources to the cosmic background from optical to radio
bands.
\item{(3)} Instead of evolving all IR sources as a single population,
they are divided into three populations (normal late-types,
starbursts, and galaxies with AGN) which are assumed to evolve
differently. Given the observational evidence (Aussel et al. 1999;
Elbaz et al. 1999; Flores et al. 1999), most of the evolution is
attributed to the starburst (interacting) galaxies.
\end{description}

In Paper II, it was found that a pure luminosity evolution of
form $L\propto (1+z)^{4.5}$ ($z \leq 1.5$) can fit the
ISOCAM 15$\mu m$ surveys and IRAS 60$\mu m$ surveys best
(both surveys are insensitive to the evolution at $z>1.5$).
This evolution rate is significantly stronger than those
derived in previous IRAS studies ($L\propto (1+z)^{3}$), while 
at same time it confirms that 
luminosity evolution models fit the data better than 
density evolution models (e.g. Pearson and Rowan-Robinson 
1996).

In this paper, while assuming that normal late-types and 
galaxies with AGNs  
undergo pure luminosity evolution ($L\propto (1+z)^{1.5}$
and $L\propto (1+z)^{3.5}$, respectively),
we found that a model assuming that both the luminosity
and the density of starburst galaxies evolve significantly
($L\propto (1+z)^{4.2}$ and $\rho \propto (1+z)^{2}$) before
the turn-over redshift (z=1.5), fits the data best.
A model with more luminosity evolution and less density
evolution ($L\propto (1+z)^{5}$ and $\rho \propto (1+z)$)
predicted a peak in the Euclidean normalized differential 
counts at 15$\mu m$ too shallow and too bright compared to the data (fig.7c).
Another model with less luminosity evolution and more density
evolution ($L\propto (1+z)^{3.5}$ and $\rho \propto (1+z)^3$),
while fitting well the 15$\mu m$ counts,
predicted too much IR background (fig.7d and Fig.8a).

The prediction of our best-fit model of a density evolution
of starburst galaxies on the order of $(1+z)^2$ is consistent 
with the theoretical prediction (Carlberg et al. 1994) and with
previous observations (Infante et al. 1996;
Roche and Eales 1999; Le F\'evre et al. 2000) of the evolution of
merger rate.  It is also very consistent with 
the close relation between starbursts and
galaxy interactions/mergers (see Sanders and Mirabel 1996
for a review).

The constraints on the evolution of IR sources at $z>1.5$ are 
set by the CIB and the 850$\mu m$ SCUBA counts. Because of the
negative K-correction, high redshift galaxies contribute significantly
to the source counts and the cosmic background in the submm bands.
It is found that in order to avoid over-predicting the CIB
and the SCUBA counts, a steep decrease after z=1.5, in the form
of $(1+z)^{-3}$, in the evolution of IR sources is required.
This is different from the result in Paper II, where the
`Flat Model' (luminosity and density remain constant after z=1.5)
can fit the CIB satisfactorily (Fig.7 of Paper II).
This is due to the difference in the submm SEDs adopted in these
two papers. In Paper II a single template for the submm SEDs is
adopted for all sources, specified by T=40K and $\beta = 1.5$
(Blain et al. 1999). In this paper, the submm SEDs are determined
using empirical correlations between IRAS fluxes and the submm fluxes.
This results in a higher submm luminosity for a given MIR 
luminosity ($L_{24\mu m}$). Since little is known about the
IR-submm SEDs of high z sources, and we cannot test 
our assumption that
the correlations between the IR and submm fluxes of local galaxies
can be applied to high z galaxies,  
the rapid decrease of the SFR predicted by our best-fit model 
is highly uncertain.

Because of the steep decrease after z=1.5, our best-fit
model predicts a redshift distribution with
a single peak at z$\sim 1$ for a survey with $f_{24\mu m}=0.055 mJy$
(5$\sigma_{conf}$).
This is very different from the double peaked distribution
predicted by the `Flat Model' (Fig.19, see also Fig.14c of Paper I).
This provides a straightforward test to distinguish these models
once the redshifts of a sample of SIRTF 24$\mu m$ sources are obtained.

\subsubsection{Comparisons with Roche and Eales (1999)}
In the multi-IR-band model of Roche and Eales (1999),
a density evolution of $\rho \propto (1+z)^2$ is also assumed.
However, the luminosity evolutionary rate of starburst galaxies
found by  Roche and Eales (1999), $L \propto (1+z)^2$ ($z<1$),
is significantly weaker than predicted by our best-fit model: 
$L \propto (1+z)^{4.2}$ ($z<1.5$).

\subsubsection{Comparisons with Dole et al. (2000)}
Compared to the other two populations (normal late-types and
galaxies with AGNs), our best-fit model predicts much
stronger evolution for z$<$1.5 in the starburst galaxies population. This
is similar to the model of Dole et al. (2000). However, they
confine their starburst component, called 'ULIRGs', to
galaxies with $> 2\times 10^{11} L_\sun$. In order to fit the
ISO counts they have to evolve the 'ULIRG' population significantly,
resulting in a 'break' in the evolved luminosity function at z=2.5
(Fig.3a of Dole et al. 2000). In Fig.22, the evolved luminosity
functions predicted by our best-fit model are plotted.
At z$\geq$1, the starburst component dominates the LF
in the whole luminosity range ($10^{7}$---$10^{12}$).
Our model does not predict a
break such as presented in the model of Dole et al. (2000).

\subsubsection{Comparisons with Blain et al. (1999)}

Blain et al. (1999) used IRAS 60$\mu m$ counts, early results of
ISO 175$\mu m$ counts, SCUBA 850$\mu m$ counts, and the CIB
to constrain several families of models for the evolution of IR sources.
Since different cosmologies are used ($\Omega_m=0.3$,
$\Omega_\Lambda=0.7$ in this work, $\Omega_m=1$,
$\Omega_\Lambda=0$ in Blain et al. 1999), it is difficult to
make a quantitative comparison. In the three panels in Fig.23, 
we plot the cosmic luminosity density evolution from our best-fit 
model. In the upper panel, the MIR luminosity 
density, calculated using the parameters of
the best-fit model, is plotted versus the time since the Big-Bang
(in fraction of the age of the Universe $t_0$). In the middle
panel, the density of bolometric 
luminosity (0.1 --- 1000 $\mu m$) obtained from a simulation
($f_{24\mu m} > 10^{-9} mJy$, 0.01 deg$^2$) based on the best-fit model
is plotted. In the lower panel, results for the density of IR
luminosity (3 --- 1000 $\mu m$) from the same simulation
 is plotted. For IR sources, the IR luminosity density
 is proportional to the star formation
rate per unit cosmic comoving volume. 

Comparing the lower panel of Fig.23 with Fig.9 of Blain
et al., our best-fit model is closer to their
'Peak-Models' (rise---peak---drop) than the 'Anvil-Models' 
(rise---peak---flat). As pointed out by Blain et al. 
(1999), if indeed the contributions from AGNs to the
CIB and to the submm counts are negligible, 
the 'Peak-Models' are favored because the 'Anvil-Models'
may over predict the metal contents of the Universe. 

\subsubsection{Comparisons with Rowan-Robinson (2001)}
Recently Rowan-Robinson (2001, hereafter RR01) developed a model to study
the evolution of galaxies using multi-band IR surveys.
This goal is similar to that which motivated this paper. Consequently,
there are many similarities between RR01 and the work presented here. 
\begin{description}
\item{(1)} Strategy: Both RR01 and we use all the
   available surveys to constrain the evolution, then use the
   best-fit model to make predictions for future surveys. 
\item{(2)} Methodology: Both RR01 and we 
use local SEDs to link all surveys together,
   though RR01 started from the IRAS 60$\mu m$ band and 
   we started from IRAS 25$\mu m$ band.
\item{(3)} Results:  Both the models of RR01 and of this paper 
can fit all ISO surveys, SCUBA 850$\mu m$ surveys, and the
CIB. 
\item{(4)} These agreements are the natural consequence of the fact that in both
 papers the model parameters are  tuned
   to fit the counts in these bands. In Fig.24a, a comparison between
the model predictions on the cosmic luminosity density evolution
by RR01 (the model for $\Lambda = 0.7$, dot-dot-dot-dashed curve)
and by our best-fit
model (`Peak Model', solid curve) are plotted.
Though the function forms are
   different, the overall trends are quite similar. 
\end{description}

On the other hand, there are some 
major differences between RR01 and this paper:
\begin{description}
\item{1)} Assumptions:
RR01 assumed that all the IR sources in his model, which he
decomposed into four spectral components (cirrus, M82-like starburst,
Arp220-like starburst, and AGN dust torus), evolve as a single
population. We have assumed that the three populations
(normal late-types, starbursts, and galaxies with AGNs)
in our model evolve differently. RR01 also adopted
a different functional form (exponential) to describe the
evolution, while the evolution functions in this paper
are simple power-laws.
\item{2)} SEDs: We have used 800+ empirically determined
SEDs, which are binned according to population and luminosity,
to constrain the K-corrections and the links between counts
in different bands. This is more sophisticated than the approach
of RR01, where four theoretical SEDs taken from the literature
(Efstathiou et al. 2000b; Yoshii and Takahara 1988)
are adopted for the four populations, respectively.
\item{3)} Evolution of SEDs: In RR01 models, the SEDs of IR sources
undergoing luminosity evolution do not evolve. This is different
from our assumption for the luminosity dependence of SEDs,
namely when the luminosity of the IR sources evolves, the SED 
changes with the luminosity (see Section 6.3).
\item{4)} Predictions for redshift distribution of SIRTF sources: In
Fig.24 we compare the redshift distribution of sources in the SIRTF
70$\mu m$ band ($f_{70\mu m} \geq$ 5 mJy) predicted by RR01 and by our
best-fit model. While RR01 predict that most of these sources
('cirrus' and starbursts) have $z < 0.5$ (median $\sim 0.4$), our
best-fit model predicts the opposite, namely most of sources
(starbursts) have $z > 0.5$ (median $\sim 1$).  This difference is
mainly due to the different evolution rates of
normal-spirals ('cirrus galaxies') in the two models: In our best-fit
model, the evolution rate of the normals is rather low (evolution
index = 1.5).  Since the normals dominate sources in the local
Universe, the evolution rate of overall IR sources at low z is low in
our best-fit model.  Only when the starbursts overtake the normals as
the dominant population, which occurs at $z\sim 0.5$, does the high
evolution rate of the starbursts start to significantly affect the IR
sources on the whole. On the other hand, in RR01, all populations
evolve with the same rate.  So the evolution index is $\sim 3$ at small
redshifts for normal spirals.  Consequently, as shown in the 
upper panel of Fig.23,
the star formation rate (roughly proportional to the luminosity
density) of RR01 is about a factor of 2 higher than our best-fit model
at $z<0.5$. The SFR predicted by our best-fit model
starts to catch up with RR01 at larger z and then overshoots until z$\sim 1.5$.  
It follows that a redshift survey of SIRTF sources
in the 70$\mu m$ band will serve as a good test to distinguish the
models.  It should be noted that for the sample of faint IRAS 60$\mu
m$ sources ($f_{60\mu m} \geq 200$ mJy, Oliver et al. 1996), our
best-fit model predicts too many (about a factor of 10) sources with
z$>0.4$ compared to the observations (6 out of 1400: Rowan-Robinson,
private communication).  However, it is not very clear to what
extent the discrepancy is due to biases introduced by large scale 
structures and
by the incompleteness of the redshift survey (see the end of Section
4). RR01, which predicts about a factor of 2 less such sources than
our best-fit model, gives a better fit to the redshift data.
\item{5)} CIB:
The difference in the evolution rates in different populations
between RR01 and our best-fit model results in a significant
discrepancy on the  relative contributions from different populations
to the counts and to the CIB.
As shown in Fig.25, our model predicts that the IR-submm CIB is
predominantly due to the starburst galaxies. 
The best-fit model of RR01 predicts that most of the
CIB is due to 'cirrus galaxies' 
(close to 'normals' in our model) which, because
they are assumed to evolve as strongly as starburst
galaxies, maintain their dominance to
the IR emission from the local Universe throughout the high-z
Universe. Both of our models predict negligible contributions from
AGNs to the CIB ($< 10 \%$ at any given wavelength), although the
`starburst' model populations could include a significant 
heavily obscured segment of the AGN population that
are thought to contribute 
significantly to the Cosmic Xray Background (XRB; eg. Gilli, Salvati 
and Hasinger 2001).
\end{description}

\subsection{Star Formation History}

Much attention has been attracted since Madau et al. (1996)
related source counts and reshift distributions obtained from
deep UV/optical surveys to the star-formation/metal-production
history of the Universe. Since then the so-called ``Madau-Diagram''
has been revised many times through various improvements, including
(1) the corrections for the effect of dust extinction
on UV/optical luminosities (Madau et al. 1998; Pozzetti et al. 1998,
Steidel et al. 1999), (2) results from less extinction sensitive
Balmer line surveys (Gallego et al. 1995;
Tresse and Maddox 1998; Yan et al. 1999; Glazebrook et al. 1999),
(3) results from submm SCUBA surveys (Barger et al. 1999). 

Adopting the conversion factor of Kennicutt (1999),
SFR ($M_\sun$ yr$^{-1}$)=$L_{IR}\times 4.5\; 10^{-44}$ (erg s$^{-1}$),
we convert the IR (3---1000$\mu m$)
luminosity density curve plotted in Fig.23 (lower panel)
to a star formation rate curve.
In Fig.26 the result is compared with the survey data found in
the literature.
Not surprisingly, the
model prediction is in very good agreement with the 
results from ISOCAM surveys (Flores et al. 1999)
because at $z < 1.5$ the model is constrained by the ISOCAM data.
On the other hand, the model prediction 
is slightly higher than the results of the
UV/optical and H$\alpha$ surveys of $z<1.5$.
After the turn-over redshift, our best-fit model predicts a quick
decrease, similar to what  have been suggested by
the UV data of Lyman-break galaxies. 

The `Flat Model', represented by the dashed line, predicts
too much CIB (Fig.9) and too many counts in the SCUBA 850$\mu m$ band 
(Fig.10). The SFR of z$\gsim 3$ galaxies
predicted by the `Flat Model' is about a factor
of 7 higher than those derived from the UV data of Lyman-break galaxies,
even after the UV data are corrected for the extinction
(about a factor of 3, Steidel et al. 1999).
Although we can not rule out the
`Flat Model' because of the 
uncertainties associated with the evolution of the submm SEDs,
and those associated with the extinction corrections of the UV
data, it is certainly disfavored by our results.
On the other hand models between the `Peak Model' and the
`Flat Model', such as the Model 5 plotted by the dotted line
(see also Fig.9), are certainly allowed by our results.

\subsection{Uncertainties Introduced by Model Assumptions}

The most important assumption in this work concerns the
applicability of the SED vs. IR luminosity relation, found for
local IR galaxies (Soifer and Neugebauer 1991; Fang et al. 1998), 
for high z galaxies. Namely we assume that the SEDs do not evolve with
time for a given luminosity. Note that this is different
from assuming that the SEDs of galaxies do not evolve at all,
because our models indeed predict strong luminosity evolution
for IR galaxies and therefore, under our assumption, more 
galaxies in the early epochs have SEDs similar to local 
luminous IR galaxies which are significantly different from the
SEDs of local L$_*$ galaxies. 

This assumption is not very well constrained.
Not much is known about the IR SEDs of high z galaxies,
in particular no current FIR ($\lambda \gsim 30\mu m$) instrument 
is sensitive enough to detect them in the FIR bands. 
A few $z\gsim 1$ galaxies have been identified
in the ISOCAM surveys (predominantly in ISOCAM 15$\mu m$ surveys,
Aussel et al. et al. 1999; Flores et al. 1999), and the 
optical SEDs of some of them
have been obtained using HST images (Flores et al. 1999).
The submm SCUBA surveys 
(Hughes et al. 1998; Barger et al. 1998; Blain et al. 1999) 
may have detected quite a few high z galaxies. 
But the
positional uncertainties of the SCUBA sources are so large 
that it is very difficult to make cross identifications in other bands.
In fact so far only a handful (3 -- 5, Frayer: private communication) 
high z SCUBA sources have published high confidence optical
identifications and redshifts  
(Ivison et al. 1998; Barger et al. 1999; Frayer et al. 1999a).
The most recent results of Ivison et
al. (2000) show that indeed the SEDs of these galaxies have similar
shapes to local ULIRGs such as Arp~220 or Mrk~231.

High-z galaxies seen in
deep surveys are high luminosity galaxies. Thus, to the extent
that they have similar
SEDs to their local counterparts, our results will be valid.
Namely, intrinsically {\it faint} galaxies at high z
may have different SEDs than
their local counterparts, but this won't have any significant effect on
the predictions for number counts and the CIB.  
It is likely that in high z ULIRGs
the luminosities in all bands are predominantly radiated in localized 
starburst regions and/or AGNs, where the physical processes
determining the luminosity and the SED (e.g. 
AGN related processes, star formation,
radiative transfer, etc.) are similar to those in local ULIRGs.
In particular, both the theoretical
arguments and observational evidence (Soifer et al. 1998;
Armus et al. 1998) show that
dust is produced rapidly after the first star formation
episode in galaxies. Therefore those high luminosity
galactic nuclei in the early Universe
are likely to be optically thick (Soifer et al. 1998;
Armus et al. 1998), similar to their local counterparts.

It is expected that for large samples of galaxies 
SEDs covering  4000\AA --- 160$\mu m$
will be available when SIRTF deep surveys and follow-up
optical/NIR surveys are carried out. Then the assumption that 
the SED vs. IR luminosity relation of
local IR galaxies also holds for high z galaxies will be
fully tested.

Throughout this paper, we have assumed the $\Lambda$ cosmology
($\Omega_\Lambda=0.7$, $\Omega_m=0.3$), which is favored by
 recent observations of
type I supernovae in distant galaxies (Perlmutter et al.
1997; Garnavich et al. 1998). Compared to the 
standard $\Omega_\Lambda=0$, $\Omega_m=1$
Einstein-de Sitter cosmology, the co-moving volume corresponding 
to a given z is significantly larger in the $\Lambda$
cosmology (Carroll et al. 1992). Althogh this volume factor
is partially balanced by relatively fainter flux for
a given luminosity and given z in the $\Lambda$
cosmology, it still affects significantly
the predictions of the SCUBA counts and the submm CIB,
where the contribution from high z sources is large.
This is one of the reasons why a steep decrease after z=1.5
is favored by our best-fit 
model (Fig.26). If the Einstein-de Sitter cosmology
were adopted, a flatter SFR at early epochs would have been
obtained.

The assumptions for the evolution rates 
of normal late-type galaxies and of
galaxies with AGNs (Section 3.3) are based on the results of
optical and NIR surveys of these sources. Whether they also 
apply to the IR bands will
have to be tested with future IR surveys
(e.g. SIRTF surveys). Nevertheless, if indeed most of the
evolution in IR bright galaxies is due to starburst galaxies (and
possibly also highly obscured AGN), as suggested
by ISOCAM surveys (Elbaz et al. 1999; Flores et al. 1999),
these uncertainties will have minimal effects on our results 
for the overall IR evolution. 

Finally, the prominent peak at $z=1.5$
in the SFR v.s. redshift plot (Fig.25) is an artifact
due to the simple two-step power-law function form adopted for the
evolution functions (Eq(7) and Eq(8)). However
the SFR at z$<$1.5 is mostly constrained 
by the ISOCAM 15$\mu m$ surveys, which indeed reach as deep as z=1.5.
As was argued in Paper II, a very strong evolution all the way
back to z${\sim}1.5$ is needed to explain the sharp peak 
at $f_{15\mu m}\sim 0.4$ mJy in the  Euclidean normalized differential 
15$\mu m$ counts (Elbaz 1999).

\section{Conclusions}
We have developed empirical `backward-evolution' models for
multiband IR surveys. 
A new Monte-Carlo algorithm is developed for this task. It
exploits a large library consisting of realistic Spectral Energy
Distributions (SEDs) of 837 local IR galaxies (IRAS 25$\mu m$
selected) from the UV (1000{\AA}) to the radio (20cm), including
ISO-measured 3--13$\mu m$ unidentified broad features (UIBs).  The
basic assumption is that the local correlation between SEDs and
Mid-Infrared (MIR) luminosities can be applied to earlier epochs of
the Universe.  A $\Lambda$-Cosmology ($\Omega_\Lambda =0.7$,
$\Omega_m = 0.3$) has been assumed through out the paper.
By attaching an SED appropriately drawn from the SED
library to every source predicted by a given model, the algorithm
enables simultaneous comparisons with multiple surveys in a wide range 
of wavebands. 

IR galaxies are divided into three populations:
(1)  normal late type galaxies ('normals'), (2)
starburst/interacting galaxies ('starbursts'), 
and galaxies with AGNs ('AGNs'). Different cosmic evolution
is assumed for these different populations.
Parameterized (power-law) luminosity evolution functions ($F_i(z)$) 
and density evolution functions ($G_i(z)$) of the form  
\begin{eqnarray}
F_i(z) & = & (1+z)^{u_i} \;\; (z \leq z_{1}) \nonumber \\ 
     & = & (1+z)^{v_i} \;\; (z_{1} < z \leq z_{0}) \nonumber
\end{eqnarray}
\begin{eqnarray}
G_i(z) & = & (1+z)^{p_i} \;\; (z \leq z_{1}) \nonumber \\ 
     & = & (1+z)^{q_i} \;\; (z_{1} < z \leq z_{0}) \nonumber
\end{eqnarray}
are adopted, with $z_1=1.5$ and $z_0=7$. 
At $z<1.5$, for 'normals' (i=1) and 'AGNs' (i=3)
it is assumed $u_1=1.5$ $p_1=0$ (passive luminosity evolution) and
$u_3=3.5$ $p_3=0$ (evolution rate of optical QSOs), respectively.
The evolution rate of 'starbursts' (i=2) at $z < 1.5$ is determined
by fitting the ISOCAM 15$\mu m$ surveys. The best fit
results are $u_2=4.2$ $p_2=2$. At $1.5\leq z \leq 7$, the
evolution of IR galaxies is mostly constrained by the
submm counts and the CIB. It is found that a `Peak Model',
with the $p_i$ and $u_i$ values described above and 
$v_1=v_2=v_3=q_2=-3$, gives the best fit.
The `Flat Model', which is the same as
the `Peak Model' at $z < 1.5$ but 
$v_1=v_2=v_3=q_2=0$ at $1.5\leq z \leq 7$, over-predicts significantly the
SCUBA counts and the CIB.

Remarkably, the best-fit model (`Peak Model')
gives good fits simultaneously to all 
data (both number counts and redshift distributions)
obtained from IR surveys, including ISOCAM 15$\mu m$,
ISOPHOT 90$\mu m$, 175$\mu m$, IRAS 60$\mu m$, and
SCUBA 850$\mu m$. Predictions for contributions of
IR-bright sources to counts in other wavebands, such as
the optical B band, the NIR K band, and
the radio continuum 20cm band, are also in agreement with
the literature. This suggests that the model is robust. 

Predictions for
number counts, confusion limits, redshift distributions,
and color-color diagrams are made for multiband surveys using the
upcoming SIRTF satellite. It is found that
several SIRTF colors can be useful indicators of
galaxy populations and redshifts.

\vskip5truecm

%\clearpage

\centerline{\bf Appendix}
\oneskip

\appendix
\section{Correlations between Flux Densities in the FIR and Submm Bands}
In this appendix, 
empirical correlations between flux densities in 
the IRAS 60$\mu m$ and 100 $\mu m$ bands
and in the submm bands at 170$\mu m$, 240$\mu m$, 450$\mu m$,
850$\mu m$, and 1200$\mu m$ are studied. The results
have been applied to the SED model in order to extend the SEDs
to the submm (up to 1200$\mu m$) wavebands (Section 2).

Flux densities in the submm wavebands were collected from the literature.
In the 170$\mu m$ band (including all data between
the 160$\mu m$ and 180$\mu m$ bands), data for 29 galaxies were found
in Devereux and Young (1992), Engargiola (1991),
Klaas et al. (1997), Hippelein et al. (1996), and
Stickel et al. (1998). In the 240$\mu m$ band, 
flux densities were found for 20 galaxies (including 14 upper
limits) in Odenwald et al. (1998). In the 450$\mu m$ band,
16 detections and 7 upper limits were found in
Rigopoulou et al. (1996), Frayer et al. (1999b),
Eales et al. (1989), Dunne et al. (2000), and Alton et al. (1998).
In the 850$\mu m$ band (including also 800$\mu m$ band), 
121 flux densities (including 1 upper limit)
were found in Lisenfeld et al. (2000), Dunne et al. (2000),
Alton et al. (1998), Rigopoulou et al. (1996), Hughes et al. (1990),
Frayer et al., (1999b), Eales et al. (1989).
In the 1200$\mu m$ band (including all observations between the
1000$\mu m$ and 1300$\mu m$ bands), 62 detections and 4 upper
limits were found in Andreani and Franceschini (1996), 
Rigopoulou et al. (1996), Hughes et al. (1990),
Eales et al. (1989), Chini et al. (1986), Fich and Hodge (1993).
Color-color diagrams of $\log (f_{submm,i}/f_{100\mu m})$ versus 
$\log(f_{100\mu m}/f_{60\mu m})$, where $f_{submm,i}$ is one of the 
submm flux densities listed above, are plotted in
Fig.27 --- Fig.32. In Fig.31 and Fig.32, color-color
diagrams of $\log (f_{1200\mu m}/f_{100\mu m})$ versus 
$\log (f_{100\mu m}/f_{60\mu m})$ are plotted for the same sources,
with the sources in the list of 
Andreani and Franceschini (1996) (19 sources) being 
aperture corrected in two different ways as given in Andreani 
and Franceschini (1996), respectively. Linear relations
in the form of 
\begin{equation}
\log (f_{submm,i}/f_{100\mu m}) = A_i \; +\; B_i\; \times\;
\log (f_{100\mu m}/f_{60\mu m})
\end{equation}
were derived (eyeball) from these color-color diagrams. In Table A1, 
the resulting $A_i$ and $B_i$ are listed. 
For a given source (with detected IRAS flux densities
$f_{60\mu m}$ and  $f_{100\mu m}$)
in the SED sample, the submm flux density in
any of the given bands is then estimated using the relation:
\begin{equation}
\log (f_{submm,i}) = \log (f_{100\mu m}) + A_i \; +\; B_i\; \times\;
\log (f_{100\mu m}/f_{60\mu m}).
\end{equation}

It should be noted that the correlations plotted in the FIR/submm
color-color diagrams are generally rather poor. There are often
only a small number of sources in a plot, and many of them are
upperlimits. The data are also very heterogeneous, obtained from
observations with widely different apertures. Aperture
corrections were included only when available in the literature.
When there are more than one observations for a given source
in a given band, the data from the observation with the largest
aperture is taken. Consequently, the lines in the color-color
plots  which represent
our best (eyeball) estimates of the
$\log (f_{submm,i}/f_{100\mu m})$ versus 
$\log (f_{100\mu m}/f_{60\mu m})$ relations are very uncertain.
This reflects the true situation in the current
literature of extragalactic submm sources. Nevertheless,
it appears that the submm SEDs derived using flux densities
predicted by these relations agree well with the observations
for a wide variety of extragalactic sources (Fig.2).
And the good agreement between the simulated and measured
850$\mu m$ and 1.2mm luminosity functions (Fig.5 and Fig.6)
further support the validity of these empirical relations.
We have deliberately avoided any modelling in deriving the
relations, keeping them purely empirical. They will
set constraints to theoretical models of dust heating
in galaxies (e.g. Silva et al. 1998; Dale et al. 2001; 
Popescu et al. 2000).

%

%%%%%%%%%%%%%%%%%%%%%%%%%%%%%%%%%%%%%%%%%%%%%%%%%%%%%%%%%%%%%%%%%%%%%%%
% Acknowledgment:
%%%%%%%%%%%%%%%%%%%%%%%%%%%%%%%%%%%%%%%%%%%%%%%%%%%%%%%%%%%%%%%%%%%%%%%

\vskip2cm 
Glenn Morrison is acknowledged for help in collecting the VLA data.
George Rieke is thanked for pointing out a computational error
in a previous version of this paper. Helpful comments from
Andrew Blain, George Helou, Rob Ivison, Tom Soifer, and
an anonymous referee are acknowledged.
This research has made use of the NASA/IPAC Extragalactic Database
(NED) which is operated by the Jet Propulsion Laboratory, California
Institute of Technology, under contract with the National Aeronautics
and Space Administration.  This work has made use of data
products from the Two Micron All Sky Survey (2MASS), which is a joint
project of the University of Massachusetts and the Infrared Processing
and Analysis Center/California Institute of Technology, funded by the
National Aeronautics and Space Administration and the National Science
Foundation. This work has made use of data
services of the InfraRed Science Archive (IRSA) at the Infrared Processing
and Analysis Center/California Institute of Technology, funded by the
National Aeronautics and Space Administration. 
This work has also made use of data
products from the NRAO VLA Sky Survey (NVSS) and
the Faint Images of the Radio Sky at Twenty-cm (FIRST), both are
NRAO projects carried out using the NRAO Very Large Array (VLA).
The authors were supported by the Jet Propulsion
Laboratory, California Institute of Technology, under contract with
NASA.

%%%%%%%%%%%%%%%%%%%%%%%%%%%%%%%%%%%%%%%%%%%%%%%%%%%%%%%%%%%%%%%%%
% The references section.
%%%%%%%%%%%%%%%%%%%%%%%%%%%%%%%%%%%%%%%%%%%%%%%%%%%%%%%%%%%%%%%%%

%%%%%%%%%%%%%%%%%%%%%%%%%%%%%%%%%%%%%%%%%%%%%%%%%%%%%%%%%%%%%%%%%
% Now the tables:
%%%%%%%%%%%%%%%%%%%%%%%%%%%%%%%%%%%%%%%%%%%%%%%%%%%%%%%%%%%%%%%%%

\clearpage

\noindent{\bf Table 1. Bins of SED library}
\nopagebreak

\hskip-0.5truecm\begin{tabular}{lccc}\hline
population & \multicolumn{2}{c}{$^\dagger L_{25\mu m}$ bin} & number of \\
           & $\log (L_1)$ & $\log (L_2)$ & sources \\
\hline
Normals & 6 & 8 &  3 \\
       & 8 & 9 &  81 \\
       & 9 & 9.4 & 65  \\
       & 9.4 & 9.8 & 63  \\
       & 9.8 & 10.2 &  36 \\
       & 10.2 & 11 & 18 \\
& & & \\
Starbursts & 6 & 8 & 2 \\
       & 8 & 9 &  31 \\
       & 9 & 9.4 &  46 \\
       & 9.4 & 9.8 & 85  \\
       & 9.8 & 10.2 &  80 \\
       & 10.2 & 10.6 & 78 \\
       & 10.6 & 11& 41 \\
       & 11 &  12 & 16 \\
& & & \\
AGNs & 6 & 10 & 63  \\
       & 10 & 12 & 129 \\
\hline
\end{tabular}
\begin{description}
\item{$^{\dagger}$} L$_{25\mu m}=\nu$L$_\nu(25\mu m)$, in units of L$_\sun$.
Luminosity bins are defined by $\log (L_1) < \log (L_{25\mu m})
 \leq \log (L_2)$.
\end{description}
\vskip3truecm

% Table 3:
\clearpage
\setcounter{table}{2}
\begin{deluxetable}{lccccc}
\tablewidth{0pt}
\tablecaption{Parameters of LLFs of Different Populations
}
\tablehead{\colhead{Population} & \colhead{number} & \colhead{$\alpha$} 
 & \colhead{$\beta$}  & \colhead{ $L_*/L_\sun$}  & \colhead{C(normalization)} 
}
\startdata
    Normals     &  456 &  0.482  &    3.876  &   $5.7\times 10^9$ & 0.00035 \nl
    Starbursts  &  643 &  0.268  &    2.230  &   $7.9\times 10^9$ & 0.00066 \nl
    AGNs       &  356 &  0.336  &    1.691  &   $6.9\times 10^9$ & 0.000090 \nl
\enddata      
\end{deluxetable}
%\vskip3truecm

\clearpage

\noindent{\bf Table 4.} Evolution models. Parameters $z_0$ and
$z_1$ specify the galaxy formation time and the `turn-over' redshift,
respectively; $u_i$ and $v_i$ ($i=1,2,3$) specify the
luminosity evolution functions (Eq(7)), 
and $p_i$ and $q_i$ ($i=1,2,3$) specify the
density evolution functions (Eq(8)).

\nopagebreak

\hskip-0.5truecm\begin{tabular}{ccccccccccccccccc}\hline
%(1) & (2) & (3) &\multicolumn{4}{c}{(4)}& &\multicolumn{4}{c}{(5)} & & \multicolumn{4}{c}{(6)} \\
    &     &     & \multicolumn{4}{c}{Normals}  & & \multicolumn{4}{c}{Starbursts} & & \multicolumn{4}{c}{AGNs} \\
\cline{4-7}\cline{9-12}\cline{14-17}
model & $z_1$ & $z_0$ & $u_1$ & $v_1$ & $p_1$ & $q_1$ & & 
$u_2$ & $v_2$ & $p_2$ & $q_2$ & & $u_3$ & $v_3$ & $p_3$ & $q_3$ \\
\hline
&&&&&&&&&&&&&&&\\
Model 1$^{\dagger}$ & 1.5 & 7 & 1.5 & -3 & 0 & 0 & & 4.2 & -3 & 2 & -3 & & 3.5 & -3 &
    0 & 0 \\
&&&&&&&&&&&&&&&\\
Model 2$^{\ddagger}$ & 1.5 & 7 & 1.5 & 0 & 0 & 0 & & 4.2 & 0 & 2 & 0 & & 3.5 & 0 &
    0 & 0 \\
&&&&&&&&&&&&&&&\\
Model 3 & 1.5 & 7 & 1.5 & -3 & 0 & 0 & & 5 & -3 & 1 & -3 & & 3.5 & -3 &
    0 & 0 \\
&&&&&&&&&&&&&&&\\
Model 4 & 1.5 & 7 & 1.5 & -3 & 0 & 0 & & 3.5 & -3 & 3 & -3 & & 3.5 & -3 &
    0 & 0 \\
&&&&&&&&&&&&&&&\\
Model 5 & 1.5 & 7 & 1.5 & -1.5 & 0 & 0 & & 4.2 & -1.5 & 2 & -1.5 & & 3.5 & -1.5 &
    0 & 0 \\
&&&&&&&&&&&&&&&\\
%&&&&&&&\\
\hline
\end{tabular}
\oneskip
$^{\dagger}$ `Peak Model'.

$^{\ddagger}$ `Flat Model'.

\clearpage

\vskip5truecm

%\clearpage

\noindent{\bf Table A1. Parameters of $\log (f_{submm,i}/f_{100\mu m})$ 
vs. $\log (f_{100\mu m}/f_{60\mu m})$ relation \hfill}
\nopagebreak
\hskip-0.5truecm\begin{tabular}{lccccc}\hline
waveband &  detections &  upper limits &  $A_i$  &  $B_i$ &  Figure \\
\hline
170$\mu m$ & 29 &  & -0.3 & 1.0 & Fig.27 \\
240$\mu m$ & 6 & 14 & -0.7 & 1.36 & Fig.28 \\
450$\mu m$ & 16 & 7 & -1.58 & 1.44 & Fig.29 \\
850$\mu m$ & 121 & 1 & -2.40 & 1.45 & Fig.30 \\
1200$\mu m$ & 62 & 4 & -2.90 & 1.45 & Fig.31, Fig.32 \\
\hline
\end{tabular}


\begin{references}

\reference{Adelberger00} Adelberger, K.L, Steidel, C.C. 2000,
ApJ, 544, 218.

\reference{Alton98} Alton, P.B., Bianchi, S., Rand, R.J., Xilouris,
E.M., Davies, J.I., Trewhella, M. 1998, ApJ, 507, L125.

\reference{Andreani96} Andreani, P., Franceschini, A. 1996, MNRAS, 283, 85.

\reference{Armus98} Armus, L., Matthews, K., Neugebauer, G.,  Soifer,
B.T. 1998, ApJL, 506, 89.

\reference{Aussel99} Aussel, H., Cesarsky, C.J., Elbaz, D.,
Starck, J.L. 1999, \aap, 342, 313.

%\reference{Aussel00} Aussel, H., Coia, D., Mazzei, P., De Zotti,
%G., Franceschini, A. 2000, A\&A Suppl., 141, 257. 

%\reference{B91} Beichman, C.A., Helou, G. 1991, ApJ, 370, L1.

\reference{Barger98} Barger, A.J., Cowie, L.L., Sanders, D.B., et al.
1998, Nature, 394, 248.

\reference{Barger99} Barger, A.J., Cowie, L.L., Sanders, D.B., et al.
1999, ApJ, 518, L5.

%\reference{Bahcall99} Bahcall, N.A., Ostriker, J.P., Perlmutter, S.,
%Steinhardt, P. 1999, Science, 284, 1481.

\reference{Becker95} Becker, R.H., White, R.L., Helfand, D.J. 
1995, ApJ, 450, 559. 

\reference{Benford99} Benford, D. 1999, Ph.D. Dissertation, Caltech.

\reference{bershady98} Bershady, M.A., Lowenthal, J.D.,
 Koo, D.C. 1998, ApJ, 505, 50. 

\reference{Bertin97} Bertin, E., Dennefeld, M., Moshir, M.
1997, A\&A, 323, 685.

\reference{bicay99} Bicay, M.D., Beichman, C.A., Cutri, R.M., Madore,
B.F. 1999, ``Astrophysics with Infrared Surveys: A Prelude to SIRTF'',
ASP Conf. Series, vol. 177, ASP (San Francesco).


\reference{blain99} Blain, A.W., Smail, I., Ivison, R.J., Kneib, J.-P.
1999, MNRAS, 302, 632.

\reference{blain00} Blain, A.W., Smail, I., Ivison, R.J., Kneib, J.-P.,
2000, in  `The Hy-Redshift Universe: galaxy formation and evolution 
at high redshift', eds. A. J. Bunker
\& W. J. M. van Breughel, ASP conference vol. 193, 
ASP: San Francisco, p. 246.


\reference{Boyle88} Boyle, B.J., Shank, T., Peterson, B.A.
1988, MNRAS, 235, 935.

\reference{Carico92} Carico, D.P., Keene, J.,
Soifer, B.T., Neugebauer, G. 1992, PASP 104, 1086.

\reference{Carlberg94} Carlberg, R. G., Pritchet, C. J. \& Infante, L. 1994,
ApJ, 435, 540.

\reference{carroll92} Carroll, S.M., Press, W., Turner, E.
1992, \araa, 30, 499.

\reference{ciliegi99} Ciliegi, P., McMahon, R. G.,
 Miley, G., Gruppioni, C., et al.  1999, MNRAS, 302, 222.

\reference{Chini86} Chini, R., Kruegel, E., Kreysa, E.  1986, A\&A, 166, L8.

\reference{clavel} Clavel, J., Schulz, B., Altieri, B., Barr, P., Claes, P.,
Heras, A., Leech, K., Metcalfe, L., Salama, A. 2000, A\&A, 357, 839.

\reference{Clements99} Clements, D. L., 
Desert, F.-X., Franceschini, A., Reach, W. T.,
Baker, A. C., Davies, J. K., \& Cesarsky, C. 1999, A\&A 346, 383.

\reference{Connolly97} 
Connolly, A.J., Szalay, A.S., Dickinson, M., 
SubbaRao, M.U., \& Brunner, R.J. 1997, ApJ, 486, L11.

\reference{condon84} Condon, J.J. 1984, ApJ, 284, 44.

\reference{condon98}
Condon J.J., Cotton, W.D., Greusen, E.W. et al. 1998, AJ, 115, 1693.

\reference{condon90} Condon, J.J., Helou, G., Sanders, D. B.,
Soifer, B.T. 1990, ApJS, 73, 359.

\reference{Dale01} Dale, D.A., Helou, G., Contursi, A., Silbermann, N.A.,
Sonali, K. 2001, ApJ, 549, 215.

\reference{Devereux92} Devereux, N.A., Young, J.S. 1992, AJ, 103, 1536.

\reference{Dickinson01} Dickinson, M., et al. 2001, 
http://www.stsci.edu/science/goods. 

\reference{dol00} Dole, H., Lagache, G., Puget, J-L., et al. 2000,
astro-ph/0002283.

\reference{dol01} Dole, H., Gispert, R., Lagache, G., Puget, J-L., et al. 2001,
A\&A, 372, 364.

\reference{driver95} Driver, S.P., Windhorst, R.A., Osterander, E.J.,
et al. 1995, ApJ, 449, L23.

\reference{Dunne00} 
Dunne, L., Eales, S., Edmunds, M., Ivison, R., Alexander, P.,
Clements, D.L. 2000, MNRAS, 315, 115.

\reference{dwe98a} Dwek, E., Arendt, R.G. 1998, ApJ, 508, L9.

\reference{dwe98} Dwek, E., Arendt, R.G., Hauser, M.G., et al. 1998,
ApJ, 508, 106.

\reference{dwe98} Dwek, E., Slavin, J. 1994, ApJ, 436, 696.

\reference{ealea89} Eales, S.A., Wynn-Williams, C.G., Duncan, W.D.
 1989, ApJ 339, 859.

\reference{efstathiou00a} Efstathiou, A., Oliver, S., Rowan-Robinson,
M., Surace, C., et al. 2000a, MNRAS, 319, 1169.

\reference{efstathiou00b} Efstathiou, A.,  Rowan-Robinson,
Siebenmorgen, R. 2000b, MNRAS, 313, 734.

\reference{elb98a} Elbaz, D., Aussel, H., Baker, A.C., et al. 1998a, 
The Next Generation Space Telescope: Science 
Drivers and Technological Challenges, 34th 
Liege Astrophysics Colloquium, p. 47.

\reference{elb98b} Elbaz, D., Aussel, H., Cesarsky, C.J., et al. 1998b, 
in {\it The Universe as Seen by ISO}, eds. P. Cox \& M.F. Kessler,
p999.

\reference{elb99} Elbaz, D., Cesarsky, C.J., Fadda, D. , et al. 1999,
\aap, 351, L37.

\reference{Engargiola91} Engargiola, G. 1991, ApJS, 76, 875

\reference{Fang98} Fang, F., Shupe, D. L., Xu, C., and Hacking, P. B. 
1998, \apj, 500, 693.

\reference{Fich 93} Fich, M. Hodge, P. 1993, AJ, 415, 75.

\reference{Fink00} Finkbeiner, D.P., Davis, M., Schlegel, D.J. 2000, ApJ,
544, 81.

\reference{fix98} Fixsen, D.J., Dwek, E., Mather, J.C., Bennett,
C.L., Shafer, R.A. 1998, ApJ, 508, 123.

\reference{flores99} Flores, H., Hammer, F., Thuan, T.X.,
Cesarsky, C., Desert, F.X., et al. 1999, ApJ, 517, 148.

\reference{fra98} Franceschini, A., Andreani, P., Danese, G. 1998,
MNRAS, 296, 709.

\reference{fra88} Franceschini, A., Danese, L., De Zotti, G.,, Xu, C. 1988,
MNRAS, 233, 175.

\reference{Frayer99a} Frayer D.T., Ivison, R.J., Scoville, N.Z.,
et al. 1999a, ApJ, 514, L13.

\reference{Frayer99b} Frayer D.T., Ivison, R.J., Smail, I., Yun, M.S.
Armus, L. 1999b, AJ 118, 139.

\reference{gallego} Gallego, J., Zamorano, J.,
 Aragon-Salamanca, A., Rego, M.  1995, ApJ, 455, L1.

\reference{gardner96} Gardner, J. P.,
 Sharples, R. M., Carrasco, B. E.. Frenk, C. S.
 1996, MNRAS, 282, L1.

\reference{gardner97} Gardner, J. P.,
 Sharples, R. M., Carrasco, B. E.. Frenk, C. S.
 1997, ApJ, 480, L99.

\reference{gar98} Garnavich, P.M., Jha, S., Challis, P. et al. 1998,
ApJ, 509, 74.

\reference{gau92} Gautier, T.N., III, Boulanger, F., Perault, M.,
Puget, J.L. 1992, AJ, 103, 1313.

\reference {gill01} Gilli, R., Salvati, M. and Hasinger, G. 2001, A\&A, 
366, 407.

\reference{Glazebrook99} Glazebrook, K.,
 Blake, C., Economou, F., Lilly, S., Colless, M. 1999, 
 MNRAS, 306, 843.

\reference{Glazebrook95} Glazebrook, K., Ellis, R., Santiago, B.,
Griffiths, R. 1995, MNRAS, 275, L19.

\reference{gorjian00} Gorjian, V., Wright, E.L., Chary, R.R.
ApJ, 536, 550.

\reference{gre95} Gregorich, D.T., Neugebauer, G., Soifer, B.T., Gunn, J.E.,
Herter, T.L. 1995, AJ, 110, 259.

\reference{gri85} de Grijp, M.H.K., Miley, K.K., Lub, J., and de Jong, T.
1985, Nature, 314, 240.

\reference{hac87} Hacking, P. B., Condon, J. J., and Houck, J. R. 1987,
\apj, 316, L15.

\reference{HH87} Hacking, P., Houck, J.R. 1987, ApJS, 63, 311.

\reference{hs91} Hacking, P.B., Soifer, B.T. 1991, ApJL, 367, 49.

\reference{Hau98} Hauser, M. G., Arendt, R. G., Kelsall, T., et al. 1998,
\apj, 508, 25.

\reference{hel86} Helou, G. 1986, \apj, 311, L33.

\reference{hel90} Helou, G. Beichman, C,A. 1990, 
in {\it From Ground-Based to Space-Borne Sub-mm Astronomy},
Proc. 29th Li\`ege International Astrophysical Colloquium,
ESA SP-314, p117.

\reference{hel00} Helou, G., Lu, N.Y., Werner, M.W., Malhotra, S.,
Silbermann, N., 2000, ApJ, 532, L21.

\reference{Hippelein96}  Hippelein, H., Lemke, D., Tuffs, R.J., 
Hass, M., et al. 1996, A\&A 315, L79.

\reference{huang98} Huang, J.-S., Cowie, L., Luppino, G.
1998, ApJ, 396, 31.

\reference{Hughes90} Hughes, D.H., Gear, W.K., Robson, E.I. 
1990, MNRAS 244, 759.

\reference{Hughes98} Hughes, D.H., Serjeant, S., Dunlop, J., et al. 
1998, Nature, 394, 241.

\reference{infante} Infante, L., de Mello, Du F.,
Menanteau, F. 1996, ApJ, 469, L85. 

\reference{ivison98} Ivison, R., Smail, I., Le Borgne, J.-F., et al. 
1998, MNRAS, 298, 583.

\reference{ivison00} Ivison, R., Smail, I., Barger, A.J., et al. 
2000, MNRAS, 315, 209.

\reference{kaw1998} Kawara, K., Sato, Y., Matsuhara, H., et al. 1998,
\aap, 336, L9.

\reference{kes96} Kessler, M.F., Steinz, J.A., Anderegg, M.E.
et al. 1996, \aap, {\bf 315}, L27.

\reference{kennicutt} Kennicutt, R. C., 1999, ApJ, 525, 1165

\reference{Klaas97} 
Klaas, U., Haas M., Heinrichsen I., Schulz B. 1997, A\&A, 325, L21.

\reference{franca00} La Franca, F., Matute, I., Gruppioni, C., Alexander, D.
et al. 2000, to appear in the proceedings of the Fourth Italian
Conference on AGNs (MemSAIt) (astro-ph/0006177).

\reference{lag98} Lagache, G., Albergel, A., Boulanger, F., Puget, J.-L.
Puget 1998, \aap, 333, 709.

\reference{leFevre00} Le F\'evre, O., Abraham, R., Lilly, S.J., Ellis,
R.S., et al. 2000, MNRAS, 311, 565.

\reference{lil96} Lilly, S.J., Le F\'evre, O., Hammer, F., Crampton, D. 
1996, ApJ, 460, L1.

\reference{Lisenfeld00} Lisenfeld, U., Isaak, K.G., Hills, R. 2000, 
MNRAS, 312, 433.

\reference{lon00} Lonsdale, C.J. 2000, 
in {\it Astrophysics with Infrared Surveys: A Prelude to SIRTF},
ASPC Series {\bf 177}, eds. M.D. Bicay, C.A. Beichman, R.M. Cutri, and
B.F. Madore, p24.

\reference{lon89} Lonsdale, C.J., Hacking, P.B. 1989, ApJ, 339, L712.

\reference{lon90} Lonsdale, C.J., Hacking, P.B., Conrow, T.B.,
Rowan-Robinson, M. 1990, ApJ, 358, 20.

\reference{lon01} Lonsdale, C.J., et al.  2001,
http//www.ipac.caltech.edu/SWIRE/

\reference{Mad96} Madau, P., Ferguson, H.C., Dickinson, M., Giavalisco, M.,
et al. 1996, MNRAS, 283, 1388.

\reference{Mad98} Madau, P., Pozzetti, L., Dickinson, M. 1998,
ApJ, 498, 106.

\reference{m98} Malkan, M.A., Stecker, F.W. 1998, ApJ, 496, 13.

\reference{metcalfe91}  Metcalfe, N., Shanks, T.,
 Fong, R., Jones, L.R. 1991, MNRAS, 249, 498.

\reference{metcalfe95}  Metcalfe, N., Shanks, T.,
 Fong, R., Roche, N. 1995, MNRAS, 273, 257.


\reference{minezaki98} Minezaki, T., Kobayashi, Y.,
 Yoshii, Y., Peterson, B.A.  1998, ApJ, 494, 111.

\reference{niklas95} Niklas, S., Klein, U., Braine, J., Wielebinski, R. 1995,
A\&AS, 114, 21.

\reference{niklas97} Niklas, S., Klein, U., Wielebinski, R. 1997,
\aap, 322, 19.

\reference{Odenwald98} Odenwald, S. Newmark, J., Smoot, G. 1998, ApJ 500, 554.

\reference{Oliver96} Oliver, S.J., Rowan-Robinson, M., Broadhurst, T.J.,
et al. 1996, MNRAS, 280, 673.

\reference{Oliver00} Oliver, S., Rowan-Robinson, M., Alexander, D.M.,
et al. 2000, MNRAS, 316, 749.

\reference{pear96} Pearson, C., and Rowan-Robinson, M. 1996, \mnras,
283, 174.

\reference{pei95} Pei, Y.C. 1995, ApJ, 438, 623.

\reference{pei95} Perlmutter, S., Gabi, S., Goldhaber, G., et al.
 1997, ApJ, 483, 565.

\reference{pop00} Popescu, C.C., Misiriotis, A., Kylafis, N.D., 
Tuffs, R.J., Fischera, J. 2000, A\&A, 362, 138.

\reference{poz98} Pozzetti, L., Madau, P., Zamorani, G.,
Ferguson, H.C., Bruzual, G. 1998, MNRAS, 298, 1133.

\reference{pug96} Puget, J-L., Albergel, A., Boulanger, F., et al. 1996
\aap, 308, L5.

\reference{pug99} Puget, J-L., Lagache, G., Clements, D.L, et al. 1999,
\aap, 345, 29.

\reference{pug89} Puget, J.L., L\'eger, A.,1989, \araa, 27, 161.

\reference{Richards99} Richards, E. A., Fomalont, E. B.,
 Kellermann, K. I., Windhorst, R. A.,
 Partridge, R. B., Cowie, L. L., Barger, A. J.
1999, ApJ, 526, 73.

\reference{rigopoulou96} Rigopoulou, D., Lawrence, A., and
Rowan-Robinson, M.  1996, MNRAS, 278, 1049.

\reference{roche93} Roche, P. F., Chandler, C. J. 1993, MNRAS 265, 486.

\reference{roche99} Roche, N., Eales, S.A. 1999, MNRAS, 307, 111.

\reference{row01} Rowan-Robinson, M. 2001, ApJ, in press
(astro-ph/0012022, RR01).

\reference{row90} Rowan-Robinson, M., Hughes, J., Vedi, K., Walker, D.W.
1990, MNRAS, 246, 473.

\reference{row97} Rowan-Robinson, M., Mann, R.G., Oliver, S.J., et
al. 1997, MNRAS, 289, 490.

\reference{Rush93} Rush, B., Malkan, M.A., Spinoglio, L. 1993, ApJS, 89, 1. 

\reference{sanders96} Sanders, D.B., Mirabel, I.F. 1996, \araa, 34, 749.

\reference{sau91} Saunders, W., Rowan-Robinson, M., Lawrence, A.,
et al. 1990, MNRAS, 242, 318.

\reference{serjeant00} 
 Serjeant, S., Oliver, S., Rowan-Robinson, M., Crockett, H., et al. 2000,
MNRAS, 316, 768.

\reference{shp98} Shupe, D.L., Fan, F., Hacking, P.B., Huchra, J.P.
1998, ApJ, 501, 597.

\reference{silva99} Silva, L., Granato, G.L., Bressan, A., 
Danese, L. 1998, ApJ, 509, 103.

\reference{simpson99} Simpson, C., Eisenhardt, P. 1999, PASP, 111, 691.

\reference{soif91} Soifer, B.T., Neugebauer, G. 1991, AJ, 101, 354.

\reference{soif94} Soifer, B.T., Matthews, K., Djorgovski,
S., Larkin, J., et al. 1994, ApJL, 420, L1.

\reference{soif98} Soifer, B.T., Neugebauer, Frank, G. M., Matthews,
K., and Illingworth,  G. D. 1998, ApJL, 501, 171.

\reference{spinoglio95} Spinoglio, L., Malkan, M.A., Rush, B., 
Carrasco, L., Recillas-Cruz, E. 1995, ApJ, 453, 616.

\reference{spoon00} Spoon, H.W.W., Koornneef, J., Moorwood, A.F.M.,
Lutz, D., Tielens, A.G.G.M. 2000, A\&A, 357, 898.

\reference{Stickel98} Stickel, M., Bogun, S., Lemke, D., Klaas, U.,
et al. 1998, A\&A, 336, 116.

\reference{Stanev98} Stanev, T.,  Franceschini, A. 1998, ApJ, 494, L159.

\reference{Steidel99} Steidel, C.C., Adelberger, K., Giavalisco, M., Dickinson,
M., Pettini, M., 1999, \apj, 519, 1

\reference{trentham99} Trentham, N., Blain, A.W., Goldader, J.
1998, MNRAS, 305, 61.

\reference{Tresse} Tresse, L., Maddox, S. J. 1998, ApJ, 495, 691.

\reference{Yan} Yan, L., McCarthy, P. J., Freudling, W., Teplitz, Harry I., 
Malumuth, E. M., Weymann, . J., Malkan, M. A. 1999, ApJ, 519, L47.

\reference{will96} Williams, R.E., Blacker, B., Dickinson, M., 
Van Dyke, W., et al. 1996, \aj, 112, 1335.

\reference{Xu98} Xu, C., Hacking, P.B., Fan, F., Shupe, D.L., Lonsdale, C.J.,
Lu, N.Y., Helou, G.X., 1998, ApJ, 508, 576 (Paper I).

\reference{Xu00} Xu, C., 2000, \apj, 541, 134 (Paper II).

\reference{Yahil91} Yahil, A., Strauss, M., Davis, M., Huchra, J.P. 
1991, \apj, 372, 380.

\reference{Yoshii99} Yoshii, Y., Takahara, F. 1988, \apj, 326, 1.

\end{references}
\end{document}